\documentclass[11pt]{article}
\pdfoutput=1
\usepackage{jheppub}

\usepackage{amssymb,amsmath}
\usepackage{amsfonts}
\usepackage{mathtools}
\allowdisplaybreaks

\usepackage{physics}
\usepackage{cancel}
\usepackage{enumitem}
\usepackage[usenames,dvipsnames]{xcolor}
\usepackage{bm}
\usepackage{bbm}
\usepackage{slashed}
\usepackage{mathrsfs}

\let\originalleft\left
\let\originalright\right
\renewcommand{\left}{\mathopen{}\mathclose\bgroup\originalleft}
\renewcommand{\right}{\aftergroup\egroup\originalright}

\makeatletter
\newenvironment{equations}[1][]{\subequations\ifx\relax#1\relax\else\label{#1}\fi\align\ignorespaces}{\endalign\ignorespacesafterend\endsubequations}
\def\@spliteq#1{\begin{equation}\begin{split}#1\end{split}\end{equation}}
\def\splitequation{\collect@body\@spliteq}

\makeatother

\newcommand{\diff}{\mathrm{d}}
\newcommand{\de}{\partial}
\newcommand{\vepsilon}{\varepsilon}

\newcommand{\vphi}{\varphi}
\newcommand{\vrho}{\varrho}

\newcommand{\hi}{\bar{\imath}}
\newcommand{\hj}{\bar{\jmath}}
\newcommand{\hk}{\bar{k}}
\newcommand{\hI}{\bar{I}}
\newcommand{\hJ}{\bar{J}}

\newcommand{\hz}{\bar{z}}

\newcommand{\hM}{\overline{\mathcal{M}}}
\newcommand{\hZ}{\bar{Z}}
\newcommand{\hmu}{\bar{\mu}}
\newcommand{\hde}{\overline{\partial}}

\makeatletter
\@addtoreset{equation}{section}

\makeatother

\newcommand{\nn}{\nonumber}

\makeatletter
\gdef\@fpheader{}
\makeatother

\title{Kodaira-Spencer Anomalies with Stora-Zumino Method}

\author[a,b]{Davide Rovere}

\affiliation[a]{Dipartimento di Fisica e Astronomia ``Galileo Galilei'', Universit\`a di Padova, Via F. Marzolo 8, 35131 Padua, Italy}
\affiliation[b]{INFN, Sezione di Padova, Padua, Italy}

\emailAdd{davide.rovere@studenti.unipd.it}

\abstract{Holomorphic diffeomorphism anomalies of $2\,n$-dimensional gravitational theories in Beltrami parametrisation (Kodaira-Spencer anomalies) are computed in the \textsc{brst} framework, using an extension of the Stora-Zumino method. This method, which allows to compute anomalies in a very concise way, makes manifest the topological origin of anomalies. They have a clear geometric interpretation, since they are expressed in terms of Chern polynomials and Pontryagin invariants. The key ingredient is the formulation of the \textsc{brst} transformations in terms of polyforms, whose total degree is the sum of the form degree and of the ghost number. This approach simplifies significantly the analysis available in literature and it allows to compute many other solutions. Namely, an anomaly, which was computed using different methods, is proved to be a consistent \textsc{brst} anomaly, thereby
supplementing a conclusion in a previous analysis.}

\begin{document}
\maketitle

\section{Introduction}

In this paper we compute holomorphic or chiral diffeomorphism anomalies in $2\,n$-dimensional theories in Beltrami parametrisation, using an extension of the Stora-Zumino method. 

Stora and Zumino discovered long ago a simple method in computing anomalies, which has the advantage to make clear the geometric nature of anomalies. \cite{Stora:1976LM, Stora:1976kd, Stora:1984, Zumino:1983ew, Manes:1985df} The method, originally thought to apply to Yang-Mills theory, but soon extended to gravitational anomalies \cite{Langouche:1984gn} and to rigid supersymmetry in superspace formulation \cite{Girardi:1985hf}, provides a way to find a cohomological solution of the \emph{Stora-Zumino equation}
\begin{equation}\label{SZeq}
\delta\,\omega_{d+1} = 0,
\end{equation}
which is equivalent to the well-known anomaly descent. $\delta = \diff + s$ is a nilpotent operator, defined by means of the exterior differential $\diff$ and of the \textsc{brst} operator $s$. $\omega_{d+1}$ is a \emph{polyform} or \emph{generalised form} of total degree $d+1$, $d$ being the number of spacetime dimensions. A polyform is a formal linear combination of differential forms in the bi-complex defined by $\diff$ and $s$, each of them having the same sum between the $\diff$ grading (form degree) and the $s$ grading (ghost number). The total degree of a polyform is the degree induced by $\delta$. If $\omega_{d+1}$ is a $\delta$-cocycle, then its component $\omega_1^{(d)}$ of form degree $d$ and ghost number one is an anomaly.

The Stora-Zumino method relies on the definition of a \emph{polyconnection} $\mathcal{A}$, of total degree one, which encompasses both the usual Yang-Mills one-form connection and the ghost, and of the corresponding \emph{polycurvature}, defined by means of $\delta$ as
\begin{equation}
\mathcal{F} = \delta\,\mathcal{A} + \mathcal{A}^2.
\end{equation}
By definition, $\mathcal{F}$ has total degree two, so in general it contains three components, a two-form of ghost number zero, which is the usual curvature $F=\diff A + A^2$, a one-form of ghost number one and a zero-form of ghost number two. Nevertheless, in the Yang-Mills case, the last two components are vanishing
\begin{equation}\label{F0ghost}
\mathcal{F} = F
\end{equation}
(\emph{horizontality condition}) \cite{Baulieu:1981sb, Manes:1985df}. So, the Pontryagin invariant $\Tr\mathcal{F}^{n+1}$ vanishes too, in $2\,n$ dimensions. Therefore, the Chern polynomials $Q_{2n+1}(\mathcal{A},\mathcal{F})$, which satisfies $\delta\,Q_{2n+1}(\mathcal{A},\mathcal{F})=\Tr\mathcal{F}^{n+1}$, solves the equation \eqref{SZeq}, and so its component of ghost number one is an anomaly. 

The crucial fact in the Stora-Zumino method in $2\,n$ dimensions is not so much the horizontality of the polycurvature, but rather the possibility of defining invariants of degree $2\,n+2$ that are zero in $2\,n$ dimensions, \emph{even if} the polycurvature is not horizontal. By the way, there are many interesting cases in which polycurvatures are not horizontal. The first situation one can consider is when the horizontality is spoiled by a ghost number one component
\begin{equation}\label{F1ghost}
\mathcal{F} = F + \lambda,
\end{equation}
where $\lambda$ is a one-form of ghost number one. This is the case of conformal gravity \cite{Imbimbo:Unpublished2,Rovere:2024nwc} and of Poincaré \cite{Frob:2021sao} or conformal supergravity \cite{Imbimbo:2023sph}. In both the cases, it is possible to find invariants out of the non-horizontal polycurvatures, which turn out to be vanishing when the number of spacetime dimensions is considered.

In this paper we show that it is possible to extend the Stora-Zumino method even if the polycurvature encompasses a zero-form component $\vphi$ of ghost number two
\begin{equation}\label{F2ghosts}
\mathcal{F} = F + \lambda + \vphi.
\end{equation}
This is the case of \emph{Kodaira-Spencer theory of gravity}, which describes the closed string sector of model \textsc{b} topological string theory \cite{Witten:1992fb, Bershadsky:1993cx, Bershadsky:1993ta, Bershadsky:1994sr, Labastida:1994ss, Becchi:1997jg, Giusto:2012jm}. The celebrated Kodaira-Spencer deformation theory of complex structures relies on the Beltrami parametrisation of the metric tensor. In two dimensions this parametrisation reads 
\begin{equation}
\diff s^2 = \vrho\,|\diff z + \mu\,\diff\hz|^2,
\end{equation}
where $z,\hz$ are complex coordinates, $\vrho$ is a conformal factor and $\mu$, called \emph{Beltrami differential}, plays a crucial r\^ole in two-dimensional conformal field theories, since it encompasses the Weyl invariant part of the metric. A remarkable application is the covariant quantisation of the bosonic string \cite{Baulieu:1986hw, Baulieu:1987jz, Stora:1987kc, Lazzarini:1988rv, Lazzarini:1989gp, Stora:1990py}.

The \textsc{brst} cohomology in two-dimensional gravity in Beltrami parametrisation is computed in \cite{Bandelloni:1993cq, Bandelloni:1994vy, Brandt:1995gu}. The main result is that a diffeomorphism anomaly exists, known as \emph{Gelfand-Fuchs anomaly}
\begin{equation}
\de\,c\,\de^2\,\mu\,\diff z\,\diff\hz,
\end{equation}
where $c$ is the holomorphic (also called chiral) diffeomorphism ghost \cite{Becchi:1987as}. The Gelfand-Fuchs anomaly is equivalent to the better known trace anomaly in two dimensions (proportional to the Ricci scalar) \cite{Knecht:1990wb, Knecht:1991cr, Knecht:1991vs}. In the latter case, the diffeomorphism invariance is preserved and the Weyl invariance is broken; in the former, the Weyl invariance is preserved and the diffeomorphism invariance is broken. 

Similarly, the Beltrami parametrisation in $2\,n$ dimensions reads
\begin{equation}
\diff s^2 = \vrho_{i\hj}\,(\diff z^i + \mu^i_{\hk}\,\diff z^{\hk})\,(\diff z^{\hj} + \mu^{\hj}_k\,\diff z^k).
\end{equation}
$\mu^i_{\hi}$ is the $2\,n$-dimensional Beltrami differential. The chiral ghost becomes an anticommuting vector field $c^i$ in the holomorphic bundle. 

The computation of holomorphic anomalies in gravitational theories in this parametrisation is less studied than the two-dimensional case. A nilpotent \textsc{brst} operator on the chiral sector, given by $\mu^i_{\hi}$ and $c^i$, can be defined:
\begin{equations}
s\,c^i &= c^j\,\de_j\,c^i,\label{BeltramiBRSTRules1}\\
s\,\mu^i_{\hi} &= \de_{\hi}\,c^i + c^j\,\de_j\,\mu^i_{\hi}-\de_j\,c^i\,\mu^j_{\hi}.\label{BeltramiBRSTRules2}
\end{equations}
We will call the diffeomorphism anomalies in this sector \emph{Kodaira-Spencer anomalies}. In \cite{Losev:1996up} some partial results in computing them are presented. In \cite{Bandelloni:1998vp} there is a first (and to our best knowledge unique) systematic analysis of the \textsc{brst} cohomology in the Beltrami parametrisation in $2\,n$ dimensions, using spectral sequences and filtration techniques. There is not a complete agreement with the results of the two papers. As pointed out by the authors, an anomaly in \cite{Losev:1996up}, which is a natural generalisation of the Gelfand-Fuchs anomaly in $2\,n$ dimensions
\begin{equation}\label{ProblematicAnomaly}
\Tr(\de\,c\,(\diff\,\de\,\mu)^n),
\end{equation}
is not captured in the \textsc{brst} analysis worked out in \cite{Bandelloni:1998vp}. 

The main new result of the present paper is to show that \eqref{ProblematicAnomaly} is a perfectly consistent \textsc{brst} anomaly. To prove this, we recast the \textsc{brst} transformations in polyforms 
\begin{equation}
\delta\,\mathcal{M}^i = \mathcal{M}^j\,\de_j\,\mathcal{M}^i, \;\;
\text{where}\;\;\mathcal{M}^i = \diff z^i + \mu^i_{\hi}\,\diff z^{\hi} + c^i,
\end{equation}
and we apply an extension of the Stora-Zumino method. The whole problem of computing anomalies in this setting can be reconsidered and the application of the Stora-Zumino method simplifies significantly the analysis in \cite{Bandelloni:1998vp}. All the anomalies computed in \cite{Bandelloni:1998vp} can be captured by the Stora-Zumino method, and many other ones are produced, whose number increases as $n$ increases. The general expression of the anomalous cocycle we computed is the following:
\begin{equation}
(\omega_{k_1k_2\dots k_p})_1^{(2n)} \simeq
\sum_{l=1}^{n+1} k_l\,\Tr(\diff\,\de\,\mu)^{k_1}\dots \Tr(\de\,c\,(\diff\,\de\,\mu)^{k_l-1})\dots\Tr(\diff\,\de\,\mu)^{k_p},
\end{equation}
where $k_1,k_2\dots k_p$ are positive integers, satisfying $k_1 + \dots + k_{p} = n+1$, with $1 \leqslant k_1 \leqslant \dots \leqslant k_{p} \leqslant n+1$, $p$ being the number of (possibly repeated) integers in the partition.

In the second section we briefly review the Stora-Zumino method and we reconsider the well-known type \textsc{a} anomaly in arbitrary even dimensions \cite{Deser:1993yx}, showing that it is a first simple example of application of the Stora-Zumino method extended to non-horizontal polycurvatures.  In the third section we review the $2\,n$ Beltrami parametrisation, which is reformulated in terms of polyforms in the fourth section. In the fifth section we show that it is possible to determine the Gelfand-Fuchs anomaly using an extension of the Stora-Zumino approach. In the sixth section we apply this extension to arbitrary even dimensions. Finally, in the conclusions, we compare with the results in \cite{Losev:1996up} and \cite{Bandelloni:1998vp}.

\section{Review and extension of the Stora-Zumino method}

Let us start by reviewing the usual Stora-Zumino method \cite{Stora:1976LM, Stora:1976kd, Stora:1984, Zumino:1983ew, Manes:1985df}. As well known, in the \textsc{brst} formalism \cite{Becchi:1974md, Becchi:1974xu, Becchi:1975nq, Tyutin:1975qk} the task of finding the possible anomalies in a quantum field theory is phrased in terms of a cohomological problem:\footnote{For an overview of the \textsc{brst} formalism applied to anomalies see for example \cite{Bertlmann:2000, Piguet:1995er} and the references therein.} In $d$ dimensions, one has to compute the cohomology of the \textsc{brst} operator $s$ on local integrated $d$-forms of ghost number one, constructed as polynomials in the space of fields and ghosts of the theory. Equivalently, one has to solve the anomaly descent which defines the $s$-cohomology on local $d$-forms of ghost number one modulo the exterior differential $\diff$. The descent reads
\begin{equations}
& s\,\omega_g^{(d+1-g)} = -\diff\omega_{g+1}^{(d-g)}, \quad g = 1,\dots d,\\
& s\,\omega_{d+1}^{(0)} = 0,
\end{equations}
$\omega_g^{(p)}$ being a $p$-form of ghost number $g$.  
Stora and Zumino observed that, since all the forms involved in the descent have the same sum between the form degree and the ghost number, the whole descent can be recasted into a single equation (\emph{Stora-Zumino equation}):
\begin{equation}\label{SZeqBis}
\delta\,\omega_{d+1} = 0,
\end{equation}
where 
\begin{equation}
\delta = \diff + s.
\end{equation}
and
\begin{equation}
\omega_{d+1} = \omega^{(d)}_1 + \dots + \omega_{d+1}^{(0)}.
\end{equation}
$\delta$ is a nilpotent operator, since $s$ and $\diff$ are nilpotent and they anticommute \cite{Dubois-Violette:1985bnc, Dubois-Violette:1985isr, Brandt:1989gv, Brandt:1989gy, Brandt:1989rd}. $\omega_{d+1}$ is an example of \emph{polyform} or \emph{generalised form}. $s$ and $\diff$ define a bi-complex, since both induce a grading on the differential forms, the ghost number and the form degree. A polyform is a linear combination of differential forms in the bi-complex with the same sum between the form degree and the ghost number, called \emph{total degree}.\footnote{A similar setting works in  equivariant localisation, which has recently attracted interest for its application in holography and supergravity (see \cite{Martelli:2023oqk, Colombo:2023fhu, BenettiGenolini:2023kxp, BenettiGenolini:2023ndb, BenettiGenolini:2024kyy}). In that case, $s$ is replaced by the contraction operator $-\iota_\gamma$ with respect to a vector field $\gamma^\mu$, whose grading counts twice the number of $\gamma$'s. In this way, $\delta$ corresponds to $\diff -\iota_\gamma =: \diff_\gamma$, known as \emph{equivariant differential}. It defines the equivariant cohomology on Lie invariant differential forms, that is, on forms $\omega$ such that $\mathcal{L}_\gamma\,\omega = \{\iota_\gamma,\diff\}\,\omega = 0$, since $\diff_\gamma^2 = -\mathcal{L}_\gamma$. Equivariant cohomology was introduced by H. Cartan and A. Weil, who defined two different models, whose equivalence was proved by Kalkman \cite{Kalkman:1993zp} (see also \cite{Dubois-Violette:1986vtp, Stora:1993zw}).} This means that a polyform $\Omega_n$ of total degree $n$ in $d$ dimensions can be expanded in its components $\Omega_{n-p}^{(p)}$, with form degree $p=0,\dots,n$. The total degree is the grading induced by $\delta$ on the space of polyforms. The set of polyforms equipped with $\delta$ can be treated in complete analogy with the set of ordinary differential forms with the ordinary exterior differential $\diff$. The unique crucial difference between ordinary forms and polyforms is that, whereas an ordinary form has to vanish if its form degree exceeds the number of spacetime dimensions, a polyform does not. Indeed, it may contain components with form degree less than $d$, even if the total degree exceeds $d$.

Consider the Yang-Mills one-form connection $A$ and the ghost $c$ in $2\,n$ dimensions and denote with $F=\diff A + A^2$ the corresponding curvature. Define the \emph{polyconnection}, of total degree one, and its \emph{polycurvature}, of total degree two,
\begin{equation}
\mathcal{A} = A + c, \quad \mathcal{F} = \delta\,\mathcal{A} + \mathcal{A}^2.
\end{equation}
Then, the Chern-Simons polynomial $Q_{2n+1}(\mathcal{A},\mathcal{F})$ of degree $2\,n+1$ solves the equation \eqref{SZeq} or \eqref{SZeqBis}, since it is a polyform of total degree $2\,n+1$ and 
\begin{equation}\label{PontrCS}
\Tr \mathcal{F}^{n + 1} = \delta\,Q_{2\,n+1}(\mathcal{A},\mathcal{F}),
\end{equation}
where the left-hand side is the Pontryagin invariant  out of $\mathcal{F}$, which is a polyform of total degree $2\,n+2$. Indeed, the Yang-Mills \textsc{brst} transformations are equivalent to the fact that $\mathcal{F}$ is equal to the simple curvature $F$
\begin{equations}\label{YMAlgebra}
s\,c = -c^2, \; s\,A = -\diff c - [A,c] \Leftrightarrow \mathcal{F} = F.
\end{equations}
One says the polycurvature to be \emph{horizontal} \cite{Baulieu:1981sb, Manes:1985df}.\footnote{This property is also known as ``Russian formula". As explained by Stora in \cite{Stora:1984}, the name derives from the fact that ``the formula seems to exist in the Russian literature where I was however unable to spot it".} In other terms, the unique non-vanishing component in the expansion of $\mathcal{F}$ is the component of form degree two and ghost number zero. Therefore, $\Tr \mathcal{F}^{n + 1}$ has only the component with form degree $2\,n+2$. But this vanishes in $2\,n$ dimensions. Therefore, equation \eqref{PontrCS} becomes
\begin{equation}
0 = \delta\,Q_{d+1}(\mathcal{A},\mathcal{F}),
\end{equation}
which is the equation \eqref{SZeq} or \eqref{SZeqBis}. Thus, the component $Q_1^{(d)}$ of $Q_{d+1}$ is an anomaly. 

Summarising, the properties used to get the result are:
\begin{itemize}
\item[1.] The complex of polyforms equipped with $\delta$ enjoys the same algebraic relations as the complex of ordinary differential forms with $\diff$;
\item[2.] The Yang-Mills polycurvature is horizontal.
\end{itemize}
Nevertheless, one may notice that, in order the Stora-Zumino method to work, it is sufficient to find combinations $I_{2n+2}(\mathcal{F})$ of the polycurvatures, which are of total degree $2\,n+2$ and which are $\delta$-closed (invariant)
\begin{equation}
\delta\,I_{2n+2}(\mathcal{F})=0,
\end{equation} 
but vanishing in $2\,n$ dimensions,
\begin{equation}
I_{2n+2}(\mathcal{F})=0\quad\text{in}\;2\,n\;\text{dimensions},
\end{equation}
\emph{even if} the polycurvatures are not horizontal. Expanding $I_{2n+2}(\mathcal{F})$ in components
\begin{equation}
I_{2n+2}(\mathcal{F})=I^{(2n+2)}_0 + I^{(2n+1)}_1 + I^{(2n)}_2 + \dots + I^{(0)}_{2n+2},
\end{equation} 
it is sufficient that the components $I^{(2n+2-g)}_g$ of ghost number $g>1$ vanish to get a vanishing $I_{2n+2}(\mathcal{F})$. Indeed, the components with $g=0$ and $g=1$ are automatically vanishing \emph{in} $2\,n$ \emph{dimensions}, since they are differential forms of degree $2\,n+2$ and $2\,n+1$ respectively.\footnote{Anomalies are related to the index of differential operators. An index can be usually expressed as the integral of a characteristic class, by means of family index theorems. In the non-abelian case, the link between characteristic classes and anomalies is manifest if one adds two more dimensions to the number of spacetime dimensions. This observation brings to the so-called ``two-step descent" 
procedure \cite{Zumino:1983rz}. Later on, the addition of dimensions gained an holographic interpretation \cite{Henningson:1998gx}. The integrated anomaly in a gauge theory at the boundary is equal to the \textsc{brst} variation of a Wess-Zumino-like action in the bulk: $\int_{\mathscr{M}_{d+1}} \omega^{(d+1)}_0 = - \int_{\de\mathscr{M}_{d+1}} \omega^{(d)}_1$. This can be seen by including a step more in the descent, if the spacetime dimension is increased by one: $s\,\omega_0{}^{(d+1)} = - \diff \omega^{(d)}_1$. Nevertheless, it is not necessary to extend the number of dimensions to make manifest the topological nature of anomalies and to compute them. The two-step descent approach is completely equivalent to the Stora-Zumino method \cite{Imbimbo:2023sph}.} So, we should replace the property 2 stated above with
\begin{itemize}
\item[2.] (bis) There is a $\delta$-closed polyform $I_{2n+2}(\mathcal{F})$ of total degree $2\,n+2$ which has no component of ghost number higher than one in $2\,n$ dimensions.
\end{itemize}

Notice that, $I_{2n+2}(\mathcal{F})$ must contain $n+1$ $\mathcal{F}$'s, since it has total degree $2\,n+2$, and $\mathcal{F}$ has total degree two. Expanding $I_{2n+2}$ in components $I_{g}^{(2n+2-g)}$ as the ghost number varies, one realises that: 
\begin{itemize}
\item[--] If $\mathcal{F}$ is of the form \eqref{F0ghost}, there is only $I^{(2n+2)}_{0}$, which vanishes in $2\,n$ dimensions.
\item[--] If $\mathcal{F}$ is 
of the form \eqref{F1ghost}, in general the component of maximum ghost number is $I_{n+1}^{(n+1)}$, which contains $n+1$ $\lambda$'s.
\item[--] If $\mathcal{F}$ is of the form \eqref{F2ghosts}, in general the expansion is complete, and the component with maximum ghost number $I_{2n+2}^{(0)}$ contains $n+1$ $\vphi$'s.
\end{itemize}

Recently, it was shown that in $\mathcal{N}=1$ new minimal supergravity \cite{Frob:2021sao} and $\mathcal{N}=1$ conformal supergravity in four dimensions \cite{Imbimbo:2023sph}, the Chern polynomial $Q_5(\mathcal{A},\mathcal{F})$ out of the gauge algebra is a $\delta$-cocycle. As in the Yang-Mills case, one starts from the Pontryagin invariant $P_6(\mathcal{F})$, which is a polyform for total degree six. Nevertheless, although the polycurvatures are not horizontal because of supersymmetry ghost contributions,\footnote{A supersymmetrised version of the chiral anomaly is computed in \cite{Girardi:1985hf}, using the superspace formalism. In that setting the polycurvature is horizontal since the polyconnection satisfies the Yang-Mills algebra \eqref{YMAlgebra}.} $P_6(\mathcal{F})$ contains no component of ghost number higher than one, so it vanishes in four dimensions:
\begin{equation}
0 = P_6(\mathcal{F}) = \delta\,Q_5(\mathcal{A},\mathcal{F}).
\end{equation}
and the component with form degree four of $Q_5(\mathcal{A},\mathcal{F})$ is an anomaly.

It is possible to exhibit a simpler example, without involving supersymmetry, in which the polycurvature is of the form \eqref{F1ghost}. The case is that of conformal gravity and the anomaly which can be computed extending the Stora-Zumino method is the so-called the type \textsc{a} Weyl anomaly \cite{Deser:1993yx}. In \cite{Boulanger:2007ab, Boulanger:2007st}, considering the ($\diff + s_w$)-cohomology, where $s_w$ is the \textsc{brst} operator generating the Weyl transformation, it is shown that there is a unique Weyl anomaly cocycle which solves non-trivially the descent (trivial here means that the anomaly cocycle is directly $s$-closed, so that the descent has a single equation), which is the type \textsc{a} anomaly. The result is proved by explicit evaluation. Here we want to show that it is possible to obtain it according to the Stora-Zumino paradigma.\footnote{We stress that here by Stora-Zumino paradigma we mean not only to encompass the anomaly descent in the single equation \eqref{SZeq} or \eqref{SZeqBis} (this is the problem addressed in \cite{Boulanger:2007ab, Boulanger:2007st} in conformal gravity), but also to obtain solutions of that equation, starting from an invariant which is both vanishing in $2\,n$ dimensions and $\delta$-exact.} 

Consider the \textsc{brst} operator $s$ encoding the Weyl transformation, with ghost $\sigma$, and the diffeomorphisms, with ghost $\xi^\mu$, in $2\,n$ dimensions:
\begin{equation}
s\,\xi^\mu = \tfrac{1}{2}\,\mathcal{L}_\xi\,\xi^\mu,\quad
s\,\sigma = \mathcal{L}_\xi\,\sigma,\quad
s\,g_{\mu\nu} = \mathcal{L}_\xi\,g_{\mu\nu} + 2\,\sigma\,g_{\mu\nu},
\end{equation}
where $g_{\mu\nu}$ is the metric tensor and $\mathcal{L}_\xi$ is the Lie derivative. The following identity holds \cite{Baulieu:1985md}\footnote{Notice that $s-[\iota_\xi,\diff]$ is a nilpotent operator (this follows from the $s$-transformation of $\xi^\mu$). So, $s-[\iota_\xi,\diff]$ is a well-defined \textsc{brst} operator on the space of fields, ghosts and derivatives, excluded the undifferentiated $\xi^\mu$. $[\iota_\xi,\diff]$ encompasses the universal part in the diffeomorphism transformation and it acts only on the form indices. It can be extended to the metric tensor by distinguishing a form index and a ``fiber" one (more properly, one should consider the vielbein in place of the metric tensor). In this way, $s\,(\diff x^\mu\,g_{\mu\nu}) = [\iota_\xi,\diff]\,(\diff x^\mu\,g_{\mu\nu}) + \de_\nu\,\xi^\vrho\,(\diff x^\mu\,g_{\mu\vrho}) + 2\,\sigma\,g_{\mu\nu}$. The second contribution, which should be thought as a $\text{GL}(2\,n)$ rotation (in the vielbein formulation, it is implemented by means of a Lorentz transformation), shows that $s-[\iota_\xi,\diff]$ does not correspond to the pure Weyl transformation, generated by $s_w$ defined in \cite{Boulanger:2007ab, Boulanger:2007st}. The analogue of $s-[\iota_\xi,\diff]$ in Beltrami parametrisation is $s-\xi^i\,\de_i - \xi^{\hi}\,\de_{\hi}$, which corresponds to the operator $\hat{\delta}$ defined in \cite{Bandelloni:1998vp}.}
\begin{equation}
e^{-\iota_\xi}\,(\diff + s)\,e^{\iota_\xi} = \diff + s - [\iota_\xi,\diff],
\end{equation}
where $\iota_\xi$ is the contraction operator, defined by $\iota_\xi\,\vphi=0$, for any scalar $\vphi$, and $\iota_\xi\,\diff x^\mu = \xi^\mu$, and $[\iota_\xi,\diff] = \iota_\xi\,\diff +(-)^{|\xi|}\,\diff\,\iota_\xi$, $|\xi|$ being the $\xi^\mu$ grade (equal to one in our case, since $\xi^\mu$ is anticommuting). It shows that the ($\diff + s$)-cohomology on the space of $\xi^\mu$, $\sigma$, $g_{\mu\nu}$ and their derivatives, and the ($\diff + s - [\iota_\xi,\diff]$)-cohomology on the same space, but without the undifferentiated $\xi^\mu$, are isomorphic: If $\hat{Q}$ is a ($\diff + s - [\iota_\xi,\diff]$)-cocycle, than $e^{\iota_\xi}\,\hat{Q}$ is a ($\diff + s$)-cocycle, and the mapping goes in the opposite direction too, since the exponential is invertible. Notice that, since we are considering the local cohomology,\footnote{``Local" here means that we are considering the space of polynomials in the fields and in their derivatives.} the derivatives of $\xi^\mu$ are independent on $\xi^\mu$ and they are not excluded. 

As well known, the type \textsc{a} Weyl anomaly is $\sigma\,e_{2n}(R)$, where $R = (R^\mu{}_\nu) = (\frac{1}{2}\,R_{\alpha\beta}{}^\mu{}_\nu\,\diff x^\alpha\,\diff x^\beta)$ is the matrix-valued Riemann two-form and $e_{2n}(R)$ the $2\,n$-dimensional Euler form (up to an overall factor)
\begin{equation}
e_{2n}(R) = \vepsilon^{\mu_1\nu_1\dots\mu_n\nu_n}\,R_{\mu_1\nu_1}\dots R_{\mu_n\nu_n}.
\end{equation} One can prove $\sigma\,e_{2n}(R)$ to be an anomaly by using the Stora-Zumino method in the following way. Consider
\begin{equation}\label{InvariantWeyl}
P_{2n+2} = \delta\,\sigma\,e_{2n}(\mathcal{R}),
\end{equation}
where $\delta = \diff + s - [\iota_\xi,\diff]$ and the Riemann polyform $\mathcal{R}^\mu{}_\nu$ is the curvature built with $\delta$ and the following Levi-Civita polyconnection $\mathcal{G}^\mu{}_\nu = \diff x^\vrho\,\Gamma_\vrho{}^\mu{}_\nu - \de_\nu\,\xi^\mu + \sigma\,\delta^\mu_\nu$
\begin{equation}
\mathcal{R} = \delta\,\mathcal{G} + \mathcal{G}^2.
\end{equation} 
$P_{2n+2}$ is trivially $\delta$-exact, being equal to
\begin{equation}\label{EulerAnomaly}
P_{2n+2} = \delta\,(\sigma\,e_{2n}(\mathcal{R})),
\end{equation}
since the Euler form is in turn invariant. Remarkably, although $\mathcal{R}$ is \emph{not} horizontal\footnote{It follows by considering the variation of the derivative of $\xi^\mu$, which reads $(s-[\iota_\xi,\diff])\,\de_\nu\,\xi^\mu = -\de_\vrho\,\xi^\mu\,\de_\nu\,\xi^\vrho$, and of the Levi-Civita connection $\Gamma^\mu{}_\nu = \frac{1}{2}\,\diff x^\vrho\,g^{\mu\alpha}\,(\de_\vrho\,g_{\alpha\nu}+\de_\nu\,g_{\alpha\vrho}-\de_\alpha\,g_{\vrho\nu})$, which reads $(s-[\iota_\xi,\diff])\,\Gamma^\mu{}_\nu = -\diff\,\de_\nu\,\xi^\mu - \Gamma^\mu{}_\vrho\,\de_\nu\,\xi^\vrho - \de_\vrho\,\xi^\mu\,\Gamma^\vrho{}_\nu - \diff \sigma\,\delta^\mu_\nu + \de_\nu\,\sigma\,\diff x^\mu - \de^\mu\,\sigma\,\diff x_\nu$, where notice that $\diff\sigma = \diff x^\mu\,\de_\mu\,\sigma = -\de_\mu\,\sigma\,\diff x^\mu$. In both the cases, the action of $s-[\iota_\xi,\diff]$ depends on the derivative of $\xi^\mu$, but not on the undifferentiated $\xi^\mu$.}
\begin{equation}
\mathcal{R}^\mu{}_\nu = R^\mu{}_\nu + \de_\nu\,\sigma\,\diff x^\mu - \de^\mu\,\sigma\,\diff x_\nu,
\end{equation}
the invariant $P_{2n+2}$ vanishes in $2\,n$ dimensions\footnote{See \cite{Rovere:2024nwc} for the proof, based on \cite{Imbimbo:Unpublished2}. Notice that, in this particular example, $P_{2n+2}$ starts with the component of ghost number one by the very definition \eqref{InvariantWeyl}. Indeed, we are working in the minimal set of fields in conformal gravity, without introducing a gauge field for the Weyl transformation.}
\begin{equation}
P_{2n+2}=0\quad\text{in}\;2\,n\;\text{dimensions}.
\end{equation}
Therefore, the equation \eqref{EulerAnomaly} becomes 
\begin{equation}
0 = \delta\,(\sigma\,e_{2n}(\mathcal{R})),
\end{equation}
which is the equation \eqref{SZeq} or \eqref{SZeqBis}, and so the ghost number one component $\sigma\,e_{2n}(R)$ is an anomaly.

\section{Beltrami parametrisation in $2\,n$ dimensions}

Consider a $2\,n$-dimensional Riemannian manifold with metric
\begin{equation}
\diff s^2 = g_{\mu\nu}\diff x^\mu\,\diff x^\nu
\end{equation}
and a compatible complex structure on it. The Beltrami parametrisation of the metric is
\begin{equation}
\diff s^2 = \vrho_{i\hj}\,(\diff z^i + \mu^i_{\hk}\,\diff z^{\hk})\,(\diff z^{\hj} + \mu^{\hj}_k\,\diff z^k),
\end{equation}
where $z^i, z^{\hi}$ are complex coordinates, $i,\hi = 1,\dots, n$, and $\mu^i_{\hi}$ and its complex conjugate $\mu^{\hi}_i$, known as \emph{Beltrami differentials}, parametrise the complex structure. The differential forms have a double grade $(p,\overline{p})$ according to number of $\diff z^i$ and $\diff z^{\hi}$ differentials. In $2\,n$ dimensions,  both $p$ and $\overline{p}$ have to be less than or equal to $n$ in order a $(p,\overline{p})$-form not to vanish. The Beltrami differential can be thought as a $(0,1)$-form taking values in the holomorphic vector bundle:
\begin{equation}
\mu = \mu^i\,\de_i = \mu^i_{\hi}\,\diff z^{\hi}\,\de_i.
\end{equation}

A rescaling of the coordinates $x^\mu \rightarrow e^{\sigma}\,x^\mu$, for some constant $\sigma$, can be equivalently moved on the components of the metric: $g_{\mu\nu} \rightarrow e^{2\,\sigma}\,g_{\mu\nu}$. When $\sigma$ is allowed to be local $\sigma = \sigma(x)$, one gets the Weyl scaling of the metric. The Beltrami differential should remain Weyl invariant, since $\mu^i_{\hi}\,\diff z^{\hi}$ should be homogeneous with $\diff z^i$, and $\diff z^i$, $\diff z^{\hi}$ have the same behaviour under rescaling. So, $\vrho_{i\hj}$ captures the whole part of the metric sensitive to a Weyl rescaling. 

In two dimensions (but not in higher dimensions) the Weyl scaling is equivalent to conformal transformations, so the Beltrami parametrisation of the metric is very useful to study the conformal properties of two-dimensional conformal models on curved background, as in the worldsheet picture of string theory.

The Beltrami differential $\mu$ can be thought as a deformation of  the complex structure. The Dolbeault operator $\hat{\hde} := \diff z^{\hi}\,\de_{\hi}$ becomes $\hat{\hde} - \mu$. The deformation of the complex structure is consistent if the deformed complex structure remains compatible with the Riemannian structure. In this case, there should exist a new set of complex coordinates $Z, \hZ$ (using the standard notation \cite{Lazzarini:1988rv, Lazzarini:1989gp, Stora:1987kc, Bandelloni:1998vp} in which the Dolbeault operator is the deformed one:
\begin{equation}
\de_{Z^{\hI}} \rightarrow \de_{\hi} - \mu^j_{\hi}\,\de_j.
\end{equation}
In the new complex coordinates, the metric tensor is 
\begin{equation}
\diff s^2 = \vrho_{I\hJ}\,\diff Z^I\diff Z^{\hJ},
\end{equation}
for some $\vrho_{I\hJ}$. Imposing the deformed Dolbeault operator to be nilpotent, one finds the \emph{Kodaira-Spencer equation}, which is an integrability condition for $\mu$:\footnote{Writing the deformed Dolbeault operator as $\hat{\hde} - \mu$, where $\mu$ denotes the action of the Beltrami differential induced by its bracket on vector fields, $(\hat{\hde} - \mu)^2 = \hat{\hde}^2 -[\hat{\hde},\mu] + \mu^2 = -\hat{\hde}\,\mu + \mu^2$, where we used the Leibniz identity to write $[\hat{\hde},\mu]=\hat{\hde}\,\mu$, and $\mu^2$ means $\frac{1}{2}\,[\mu,\mu]$. The Dolbeault cohomology for almost complex structures, in which the Kodaira-Spencer equation is not assumed and the Dolbeault operator is not nilpotent, is discussed in \cite{Bandelloni:1998vp, Cirici:2021}.}
\begin{equation}
\hat{\hde}\,\mu^i - \mu^j\,\de_j\,\mu^i = 0.
\end{equation}
Similarly for the complex conjugate $\hmu$. The left-hand side defines a $(0,2)$-form, called \emph{Kodaira-Spencer form}:\footnote{Note that the Kodaira-Spencer form is identically vanishing in two dimensions, since it is a $(0,2)$-form. So, in two dimensions any deformation of a complex structure remains compatible with the Riemann structure.}
\begin{equation}
F^i := \hat{\hde}\,\mu^i - \mu^j\,\de_j\,\mu^i,
\end{equation}
which can be thought as a curvature associated to $\mu$.
The Kodaira-Spencer equation is a necessary condition for the existence of the coordinates $Z, \hZ$. Actually, if one assumes that $\det(1-\mu\,\hmu)\neq 0$, it is also a sufficient condition, by means of the \emph{Newlander-Nirenberg theorem}.\footnote{According to the usual formulation of the theorem, the Kodaira-Spencer equation corresponds to the vanishing of the Nijenhuis tensor \cite{Kodaira:1986}.}

The new set of complex coordinates can be defined by  
\begin{equation}
\diff Z^I = \lambda_i{}^I\,(\diff z^i + \mu^i_{\hi}\,\diff z^{\hi}), \quad
\diff Z^{\hI} = \lambda_{\hi}{}^{\hI}\,(\diff z^{\hi} + \mu^{\hi}_j\,\diff z^j),
\end{equation}
for some integrating matrix $\lambda_i{}^I$ and its complex conjugate $\lambda_{\hi}{}^{\hI}$, which relate $\vrho_{i\hj}$ in the $z,\hz$ coordinates and $\vrho_{I\hJ}$ in the $Z,\hZ$ ones: $ \vrho_{i\hj} = \vrho_{I\hJ}\,\lambda_i{}^I\,\lambda_{\hj}{}^{\hJ}$. The partial derivatives in the new coordinates are
\begin{equations}
\de_{Z^I} &= (\det(1-\mu\,\hmu))^{-1}\,(\lambda^{-1})_I{}^i\,(\de_i - \hmu_i^{\hi}\,\de_{\hi}),\\
\de_{Z^{\hI}} &= (\det(1-\mu\,\hmu))^{-1}\,(\lambda^{-1})_{\hI}{}^{\hi}\,(\de_{\hi} - \mu^i_{\hi}\,\de_i).
\end{equations}

Under the reparametrisation
\begin{equation}
z^i \rightarrow z^i + \xi^i, \quad
z^{\hi} \rightarrow z^{\hi} + \xi^{\hi},
\end{equation}
the new complex coordinates transform as
\begin{equation}
Z^I \rightarrow Z^I + \lambda_i{}^I\,(\xi^i + \mu^i_{\hi}\,\xi^{\hi}), \quad
Z^{\hI} \rightarrow Z^{\hI} + \lambda_{\hi}{}^{\hI}\,(\xi^{\hi} + \mu_i^{\hi}\,\xi^i).
\end{equation}
Introducing the \emph{Becchi ghosts}, reflecting the integrable complex structure given by $\mu$,\footnote{Originally introduced in two dimensions in \cite{Becchi:1987as} and extended in $2\,n$ dimensions in \cite{Bandelloni:1998vp}.}
\begin{equation}
c^i = \xi^i + \mu^i_{\hi}\,\xi^{\hi}, \quad
c^{\hi} = \xi^{\hi} + \mu^{\hi}_i\,\xi^i,
\end{equation}
the transformations read
\begin{equation}
Z^I \rightarrow Z^I + \lambda_i{}^I\,c^i, \quad
Z^{\hI} \rightarrow Z^{\hI} + \lambda_{\hi}{}^{\hI}\,c^{\hi}.
\end{equation}
If a \textsc{brst} operator $s$ is introduced, encoding diffemorphisms, we can write
\begin{equation}
s\,Z^I = \lambda_i{}^I\,c^i, \quad
s\,Z^{\hI} = \lambda_{\hi}{}^{\hI}\,c^{\hi},
\end{equation}
where $c^i$ and $c^{\hi}$ are promoted to anticommuting fields.

\section{Beltrami parametrisation: Polyform formulation}

Define the following polyform of total degree one and the $\delta$ operator:\footnote{The polyform $\mathcal{M}^i$ appeared for the first time in \cite{Baulieu:1987jy} (in the two-dimensional case only).}
\begin{equation}
\mathcal{M}^i = \diff z^i + \mu^i + c^i, \quad \delta = \diff + s = \hat{\de} + \hat{\hde} + s.
\end{equation}
The expression for $\diff Z^I$ and $s\,Z^I$ are recasted into
\begin{equation}
\delta\,Z^I = \lambda_i{}^I\,\mathcal{M}^i.
\end{equation}
Similarly for the complex conjugate. Imposing $\delta$ to be nilpotent,
\begin{equation}\label{s0}
\delta\,(\lambda_i{}^I\,\mathcal{M}^i) = 0.
\end{equation}
Filtering as the ghost number varies, one gets 
\begin{equations}
& \hat{\de}\,(\lambda_i{}^I\,\diff z^i) = 0, \label{s1a}\\
& \hat{\hde}\,(\lambda_i{}^I\,\mu^i) = 0, \label{s1b}\\
& \hat{\de}\,(\lambda_i{}^I\,\mu^i) + \hat{\hde}\,(\lambda_i{}^I\,\diff z^i) = 0,\label{s1c}\\
& s\,(\lambda_i{}^I\,\diff z^i) + \hat{\de}\,(\lambda_i{}^I\,c^i) = 0,\label{s1d}\\
& s\,(\lambda_i{}^I\,\mu^i) + \hat{\hde}\,(\lambda_i{}^I\,c^i) = 0,\label{s1e}\\
& s\,(\lambda_i{}^I\,c^i) = 0,\label{s1f}
\end{equations}
whence
\begin{equations}
& \de_{[i}\,\lambda_{j]}{}^I = 0,\label{s2a}\\
& \de_{[\hi}\,(\lambda_{i}{}^I\,\mu^i_{\hj]}) = 0,\label{s2b}\\
& \hat{\hde}\,\lambda_i{}^I = \de_i\,(\lambda_j{}^I\,\mu^j),\label{s2c}\\
& s\,\mu^i = - \hat{\hde}\,c^i + c^j\,\de_j\,\mu^i - \de_j\,c^i\,\mu^j,\label{s2d}\\
& s\,\lambda_i{}^I = \de_i\,(\lambda_j{}^I\,c^j),\label{s2e}\\
& s\,c^i = c^j\,\de_j\,c^i.\label{s2f}
\end{equations}
The integrability conditions on $\lambda_i{}^I$ \eqref{s2a}, \eqref{s2b} and \eqref{s2c} are equivalent to \eqref{s1a}, \eqref{s1b} and \eqref{s1c} respectively; the $s$-variation of $\lambda_i{}^I$ \eqref{s2e} follows from \eqref{s1d}; the $s$-variation of $c^i$ \eqref{s2f} follows from \eqref{s1f}, using \eqref{s2a} and \eqref{s2e}; the $s$-variation of $\mu^i$ \eqref{s2d} follows from \eqref{s1e}, using \eqref{s2a}, \eqref{s2c} and \eqref{s2e}. 

The main property of the $s$-variation on $c^i$, $\lambda_i{}^I$ and $\mu^i$ is the \emph{holomorphic factorisation}: The complex conjugates $c^{\hi}$, $\lambda_{\hi}{}^{\hI}$, $\mu^{\hi}$ are not involved, and the $s$-variation of the latters is simply the complex conjugate of the $s$-variation of the formers: 
\begin{equation}
s\,c^{\hi} = \overline{s\,c^i},\quad
s\,\lambda_{\hi}{}^{\hI} = \overline{s\,\lambda_i{}^I},\quad
s\,\mu^{\hi} = \overline{s\,\mu^i}.
\end{equation}
Moreover, the transformations on $c^i$ and $\mu^i_{\hi}$ define a consistent truncation, since they close without involving not only their complex conjugates, but also $\lambda_i{}^I$. 

The expressions \eqref{s2a}, \eqref{s2c} and \eqref{s2e} reconstruct a polyform equation:
\begin{align}
\delta\,\lambda_i{}^I &= (\hat{\de} + \hat{\hde} + s)\,\lambda_i{}^I = 
\hat{\de}\,\lambda_i{}^I + \de_i\,(\lambda_j{}^I \,\mu^j) + \de_i\,(\lambda_j{}^I\,c^j) = \nn\\
& = \de_i\,(\lambda_j{}^I\,(\diff z^j + \mu^j + c^j)) = \de_i\,(\lambda_j{}^I\,\mathcal{M}^j).
\end{align}
Similarly for the expressions \eqref{s2d} and \eqref{s2f}:
\begin{align}
\delta\,\mathcal{M}^i &= (\hat{\de} + \hat{\hde} + s)\,(\diff z^i + \mu^i + c^i) = \nn\\
& = \hat{\de}\,(\mu^i + c^i) + \hat{\hde}\,\mu^i + (s\,\mu^i + \hat{\hde}\,c^i) + s\,c^i = \nn\\
& = \hat{\de}\,(\mu^i + c^i) + \hat{\hde}\,\mu^i + (c^j\,\de_j\,\mu^i - \de_j\,c^i\,\mu^j) + c^j\,\de_j\,c^i = \nn\\
& = \hat{\de}\,(\mu^i + c^i) + \hat{\hde}\,\mu^i - \mu^j\,\de_j\,\mu^i + (c^j+\mu^j)\,\de_j\,(\mu^i + c^i) = \nn\\
& = \hat{\de}\,(\mu^i + c^i) + F^i - \diff z^j\,\de_j\,(\mu^i + c^i) + (\diff z^j + \mu^j + c^j)\,\de_j\,(\diff z^i + \mu^i + c^i) = \nn\\
& = F^i + \mathcal{M}^j\,\de_j\,\mathcal{M}^i.
\end{align}
Summarising,
\begin{equations}
\delta\,\lambda_i{}^I &= \de_i\,(\lambda_j{}^I\,\mathcal{M}^j),\label{s3a}\\
\delta\,\mathcal{M}^i &= F^i + \mathcal{M}^j\,\de_j\,\mathcal{M}^i.\label{s3b}
\end{equations}
In this formulation we recover the holomorphic factorisation. The consistent truncation on $c^i$ and $\mu^i_{\hi}$ is elegantly collected in a single equation \eqref{s3b}. In this last expression, we left the Kodaira-Spencer form free in order to underline that the $s$-variations on $c^i$ and $\mu^i$, which is the piece of information encoded into $\delta\,\mathcal{M}^i$, are perfectly consistent and nilpotent without assuming the Kodaira-Spencer equation. Indeed, $\delta\,\mathcal{M}^i$ does not involve the integrating factor $\lambda_i{}^I$, so it does not know anything about the coordinates $Z,\hZ$. Obviously, the Kodaira-Spencer equation has to be taken into account, if $\lambda_i{}^I$ is allowed back into the play and if we want the \textsc{brst} rules on $c^i$, $\mu^i$ and $\lambda_i{}^I$ to represent the gravitational symmetry, as pointed out in \cite{Bandelloni:1998vp}.\footnote{The following simple counting argument shows that the Kodaira-Spencer equation ensures the metric tensor to have the expected number of components, when it is expressed in the Beltrami parametrisation. The metric tensor $g_{\mu\nu}$ has $n\,(2\,n+1)$ components in $2\,n$ dimensions, since it is a symmetric tensor. In the Beltrami parametrisation, one has $n^2$ components for $\vrho_{i\hj}$ and $2\,n^2$ for $\mu^i_{\hi}$ and its complex conjugate. Thus, there is a mismatch of $3\,n^2-n\,(2\,n+1) = n\,(n-1)$ components. We may impose $\frac{n(n-1)}{2}$ constraints on $\mu^i_{\hi}$, and the same number on its complex conjugate. This is precise the number of constraints imposed by the Kodaira-Spencer equation, which, setting to zero a $(0,2)$-form, imposes $\frac{n(n-1)}{2}$ conditions (plus the same number for the complex conjugate). In \cite{Baulieu:2021sem} an alternative Beltrami parametrisation of the metric is introduced, giving directly the right number of components.} Indeed, \eqref{s3a} and \eqref{s3b} are consistent with \eqref{s0} if and only if the Kodaira-Spencer equation is assumed, together with \eqref{s2a}:
\begin{align}
\delta\,(\lambda_i{}^I\,\mathcal{M}^i) &= \delta\,\lambda_i{}^I\,\mathcal{M}^i + \lambda_i{}^I\,\delta\,\mathcal{M}^i = \nn\\
& = \de_i\,(\lambda_j{}^I\,\mathcal{M}^j)\,\mathcal{M}^i + \lambda_i{}^I\,(F^i+\mathcal{M}^j\,\de_j\,\mathcal{M}^i) = \nn\\
& = \tfrac{1}{2}\,\de_{[i}\,\lambda_{j]}{}^I\,\mathcal{M}^j\,\mathcal{M}^i + \lambda_j{}^I\,\de_i\,\mathcal{M}^j\,\mathcal{M}^i + \lambda_i{}^I\,F^i + \lambda_i{}^I\,\mathcal{M}^j\,\de_j\,\mathcal{M}^i = \nn\\
& = \tfrac{1}{2}\,\de_{[i}\,\lambda_{j]}{}^I\,\mathcal{M}^j\,\mathcal{M}^i + \lambda_i{}^I\,F^i, 
\end{align}
so that
\begin{equation}
\delta\,(\lambda_i{}^I\,\mathcal{M}^i) = 0
 \Leftrightarrow 
 \de_{[i}\,\lambda_{j]}{}^I = 0, \;\; F^i = 0.
\end{equation}

Let us focus on the truncation on the algebra defined by \eqref{s3b}. In index-free notation, it reads 
\begin{equation}\label{deltaM}
\delta\,\mathcal{M} = F + \mathcal{M}^2,
\end{equation}
where $\mathcal{M}^2 = \frac{1}{2}\,[\mathcal{M},\mathcal{M}]$, the $[\cdot,\cdot]$ denoting the (graded) Lie bracket between vector fields. It is strictly analogous to the Yang-Mills case: The Kodaira-Spencer form $F$ plays the r\^ole of the (horizontal) curvature of the polyconnection $\mathcal{M}$. So, $F$ is expected to satisfy a Bianchi identity, which corresponds to the $\delta$-variation on $F$ itself:
\begin{equation}
\delta\,F = [\mathcal{M},F].
\end{equation}
The operator $\delta - \mathcal{M}$ ($\mathcal{M}$ acting by the bracket between vector fields in the adjoint representation) can be thought as the promotion to polyform of the deformed Dolbeault operator $\hat{\hde} - \mu$.

In the $\delta$ picture, it is evident the fact that a gravitational theory, enjoying the symmetry encoded in the \textsc{brst} operator, is topological if the Kodaira-Spencer equation holds. Indeed, there is no local degree of freedom in a topological theory, so the curvature vanishes. Consistently, in two dimensions, where gravity is always topological, there is no $F$ at all.\footnote{\label{NoteCS} This is also true for the three-dimensional Chern-Simons theory, if we include the antifields \cite{Alvarez-Gaume:1981klj, Bonora:1982ve, Imbimbo:2009dy}. The \textsc{brst} rules
\begin{equation}
s\,c = -c^2,\quad
s\,A =-\text{D}\,c,\quad
s\,A^* = -F - [A^*,c],\quad
s\,c^* = - \text{D}\,A^* - [c^*,c]
\end{equation}
can be recasted in a $\delta = \diff + s$ transformation with vanishing curvature, upon defining $\mathcal{A} = c^* + A^* + A + c$ (which is a polyform of total degree one, since $A^*$ is a two-form of ghost number $-1$ and $c^*$ is a three-form of ghost number $-2$):
\begin{equation}
\delta\,\mathcal{A} = -\mathcal{A}^2.
\end{equation}}

\section{Gelfand-Fuchs anomaly with polyforms}

The well-known two-dimensional trace anomaly, proportional to the Ricci scalar, is a Weyl anomaly preserving the reparametrisation invariance. Similarly, one can study the $s$-cohomology modulo $\diff$, on the space of $c, \mu$ and their derivatives, to produce a Weyl invariant anomaly, which breaks the reparametrisation invariance. This is called \emph{Gelfand-Fuchs} anomaly \cite{Fuchs:1969, Fuchs:1970a, Fuchs:1970b}. By considering appropriate (nonlocal) actions $\Gamma$ and $\tilde{\Gamma}$, whose variations give the (integrated) trace anomaly $\mathscr{A}$ and the (integrated) Gelfand-Fuchs one (plus its complex conjugate) $\tilde{\mathscr{A}}$, it was realised that a local counterterm $\mathscr{C}$ exists, linking the two actions: 
\begin{equation}\label{2dactions}
\tilde{\Gamma} = \Gamma + \mathscr{C}.
\end{equation}
This proves that the two (integrated) anomalies are equivalent \cite{Polyakov:1981rd, Polyakov:1987zb, Knecht:1990wb, Knecht:1991cr, Knecht:1991vs}. Indeed, in the \textsc{brst} setting, equivalence between anomalies means that they belong to the same class of cohomology: Taking the $s$-variation of \eqref{2dactions}, one sees that the two anomalies differ by an $s$-exact term, given by the variation of the counterterm:
\begin{equation}\label{relation_anomalies}
\tilde{\mathscr{A}} = \mathscr{A} + s\,\mathscr{C}.
\end{equation}

In two dimensions the (anti)holomorphic index $i$ ($\hi$) takes a single value $z$ ($\hz$). To simplify the notation, we will write \emph{in this section}
\begin{equations}
& \de_z \rightarrow \de, \quad \de_{\hz} \rightarrow \hde, \quad \mathcal{M}^z \rightarrow \mathcal{M}, \quad \mathcal{M}^{\hz} \rightarrow \overline{\mathcal{M}},\nn\\
& c^z \rightarrow c, \quad c^{\hz} \rightarrow \overline{c}, \quad \mu^z_{\hz} \rightarrow \mu, \quad \mu^{\hz}_z \rightarrow \overline{\mu}.
\end{equations}

It is very easy to find the Gelfand-Fuchs anomaly using the polyform approach. We have to study the $\delta$-cohomology on polyforms of degree three to produce two-dimensional anomalies, so the Stora-Zumino equation looks like
\begin{equation}
\delta\,\omega_3 = 0.
\end{equation}
The mass dimension of the component $\omega_1^{(2)}$ should be zero to get an anomaly. So, this is true for the whole $\omega_3$. The unique fields we have available are $\mathcal{M}$ and its derivatives. Indeed, the complex conjugate $\hM$ is excluded, by means of the holomorphic factorisation. Notice that the dimension is $[\mathcal{M}]=-1$, because $[\mu]=0$ and $[c]=-1$. Moreover, $[\de] = [\hde] = -1$. Actually, we should consider only the holomorphic derivative $\de$, since the $\delta$-variation of $\mathcal{M}$, which in two dimensions reads
\begin{equation}
\delta\,\mathcal{M} = \mathcal{M}\,\de\,\mathcal{M},
\end{equation}
does not produce terms depending on the antiholomorphic $\hde$.\footnote{This na\"ive argument can be refined by noticing that the $s$-variation of $c$ and the $\delta$-variation of $\mathcal{M}$ are algebraically the same. See the next section, in which this issue is directly discussed in the $2\,n$-dimensional case.} 
Because $\omega_3$ is a polyform of total degree three, we need three $\mathcal{M}$'s. To get a vanishing mass dimension, we have to include also three derivatives $\de$. There is a unique way to arrange the derivatives, bringing to the following ansatz:
\begin{equation}
\omega_3 = \mathcal{M}\,\de\,\mathcal{M}\,\de^2\,\mathcal{M}.
\end{equation}
Indeed, the other possibilities $\mathcal{M}\,\mathcal{M}\,\de^3\,\mathcal{M}$ and $\de\,\mathcal{M}\,\de\,\mathcal{M}\,\de\,\mathcal{M}$ vanish, thanks to the anticommutativity of $\mathcal{M}$, which implies $\mathcal{M}\,\mathcal{M} = 0 = \de\,\mathcal{M}\,\de\,\mathcal{M}$. 

The ansatz turns out to be a $\delta$-cocycle:
\begin{align}
\delta\,(\mathcal{M}\,\de\,\mathcal{M}\,\de^2\,\mathcal{M}) &= \delta\,\mathcal{M}\,\de\,\mathcal{M}\,\de^2\,\mathcal{M} - \mathcal{M}\,\de\,\delta\,\mathcal{M}\,\de^2\,\mathcal{M} + \mathcal{M}\,\de\,\mathcal{M}\,\de^2\,\delta\,\mathcal{M} = \nn\\
& = (\mathcal{M}\,\de\,\mathcal{M})\,\de\,\mathcal{M}\,\de^2\,\mathcal{M} - \mathcal{M}\,(\mathcal{M}\,\de^2\,\mathcal{M})\,\de^2\,\mathcal{M}\,\nn\\
& \quad + \mathcal{M}\,\de\,\mathcal{M}\,(\de\,\mathcal{M}\,\de^2\,\mathcal{M}+\mathcal{M}\,\de^3\,\mathcal{M}) = 0,
\end{align}
where each factor separately vanishes, thanks to anticommutativity. 
The cocycle in components is 
\begin{align}
\omega_3 & = (\diff z + \diff \hz\,\mu + c)\,(\diff\hz\,\de\,\mu + \de\,c)\,(\diff \hz\,\de^2\,\mu + \de^2\,c) = \nn\\
& = \diff z\,\diff \hz\,(\de\,\mu\,\de^2\,c - \de^2\,\mu\,\de\,c) \,+\nn\\
& \quad + \diff z\,\de\,c\,\de^2\,c + \diff \hz\,(\mu\,\de\,c\,\de^2\,c - c\,\de\,\mu\,\de^2\,c + c\,\de\,c\,\de^2\,\mu) \,+\nn\\
& \quad + c\,\de\,c\,\de^2\,c.
\end{align}
Integrating by parts the top-component, we find that the anomaly is 
\begin{equation}\label{GF_Anomaly}
\omega_1^{(2)} \simeq -2\,\diff z\,\diff \hz\,\de\,c\,\de^2\,\mu,
\end{equation}
where $\simeq$ means equal modulo $\diff$. $\tilde{\mathscr{A}}$ in \eqref{relation_anomalies} is the integral of $\omega_1^{(2)}$ plus its complex conjugate.

There is another way to produce the Gelfand-Fuchs cocycle $\omega_3$, mimicking the Stora-Zumino method. Consider the following polyform of total degree two 
\begin{equation}
\mathcal{R} = \mathcal{M}\,\de^2\,\mathcal{M}.
\end{equation}
It can be seen as the ``abelian curvature" associated to the ``connection-like" $\de\,\mathcal{M}$. Indeed, taking the derivative of $\delta\,\mathcal{M}$,
\begin{equation}
\delta\,\de\,\mathcal{M} = \mathcal{R},
\end{equation}
So it satisfies the Bianchi identity
\begin{equation}
\delta\,\mathcal{R} = 0.
\end{equation}
$\mathcal{R}$ is \emph{not} horizontal, since its components of ghost number one and two do not vanish:
\begin{align}
\mathcal{R} &= (\diff z + \diff\hz\,\mu + c)\,(\diff\hz\,\de^2\,\mu + \de^2\,c) = \nn\\
& = \diff z\,\diff \hz\,\de^2\,\mu + 
(\diff z\,\de^2\,c + \diff\hz\,\mu\,\de^2\,c + c\,\diff\hz\,\de^2\,\mu)
+ c\,\de^2\,c,
\end{align}
so that the previous equation is an instance of \eqref{F2ghosts}. Nevertheless, we can consider the following invariant
\begin{equation}
\mathcal{R}^2 = \mathcal{M}\,\de^2\,\mathcal{M}\,\mathcal{M}\,\de^2\,\mathcal{M},
\end{equation}
which vanishes thanks to anticommutativity
\begin{equation}
\mathcal{R}^2 = 0.
\end{equation}
On the other hand, it can be written as a $\delta$-coboundary
\begin{equation}
\mathcal{R}^2 = \delta\,\de\,\mathcal{M}\,\mathcal{R} = \delta\,(\de\,\mathcal{M}\,\mathcal{R}).
\end{equation}
So, we find
\begin{equation}
\delta(\de\,\mathcal{M}\,\mathcal{R}) = 0,
\end{equation}
which is the Stora-Zumino equation \eqref{SZeq} or \eqref{SZeqBis}, since $\de\,\mathcal{M}\,\mathcal{R}$ is a polyform of total degree three. Up to a sign, it is equal to the Gelfand-Fuchs cocycle in \eqref{GF_Anomaly}.

In this way, we have obtained the Gelfand-Fuchs anomaly using an extension of the Stora-Zumino method, since we started from a vanishing invariant out of a \emph{non-horizontal} curvature and we wrote it as a $\delta$-coboundary. In the two-dimensional case, this is just an equivalent way of writing the unique possible ansatz in the $\delta$-cohomology; but in arbitrary even dimensions, the extension of this approach will be very powerful to produce cocycles in a very simple way. Notice also that $\de\,\mathcal{M}\,\mathcal{R}$ is the abelian Chern polynomial of degree three and $\mathcal{R}^2$ is the corresponding abelian Pontryagin invariant.

\section{Kodaira-Spencer anomalies in $2\,n$ dimensions}

Consider the $\delta$-variation of $\mathcal{M}^i$, assuming the Kodaira-Spencer equation $F^i=0$ to hold:
\begin{equation}
\delta\,\mathcal{M}^i = \mathcal{M}^j\,\de_j\,\mathcal{M}^i.
\end{equation}

Notice that the component of any polyform generated by the $\mathcal{M}$'s of maximum ghost number is the same as the whole polyform with each $\mathcal{M}$ replaced by $c$. Moreover, the $\delta$-variation of $\mathcal{M}^i$ is the same as the $s$-variation of $c^i$. This suggests a correspondence between the $\delta$-cohomology on polyforms of some total degree $N$ and the $s$-cohomology on zero-forms of ghost number $N$. 

The precise isomorphism is realised by means of the following even operator, known as \emph{homotopy operator},\footnote{In the two-dimensional case it is introduced in \cite{WerneckdeOliveira:1993ig}.}
\begin{equation}
\ell\,c^i = \diff z^i + \mu^i, \quad 
\ell\,\mu^i = \ell\, \diff z^i = \ell\, \diff z^{\hi} = 0,
\end{equation}
which sends $p$-forms of ghost number $g$ into $(p+1)$-forms of ghost number $g-1$. It acts on monomials generated by the $c$'s by replacing a $c^i$ with $\diff z^i + \mu^i$. So, $\ell^N$ vanishes on polynomials of ghost number less than $N$. Moreover, $\ell$ satisfies\footnote{The existence of such an operator is a feature of all the topological theories in the \textsc{brst} setting. For example, in the three-dimensional Chern-Simons theory, one can consider \cite{Imbimbo:2009dy} (see footnote \ref{NoteCS}):
\begin{equation}
\ell\,c= A, \quad \ell\,A=2\,A^*, \quad \ell\,A^*=3\,c^*,\quad \ell\,c^*=0.
\end{equation}}
\begin{equation}
[\ell,s]=\diff, \quad [\ell,\diff]=0.
\end{equation}
The latter property is obvious, since $\ell$ does not act on $\diff z^i$, $\diff z^{\hi}$ and on the derivatives; the former property follows by explicit evaluation, and it holds on $\mu^i$ if the Kodaira-Spencer equation is assumed:
\begin{align}
(\ell\,s-s\ell)\,c^i &= \ell\,(c^j\,\de_j\,c^i) - s\,\mu^i = \nn\\
& = (\diff z^j + \mu^j)\,\de_j\,c^i + c^j\,\de_j\,\mu^i + \hat{\hde}\,c^i - c^j\,\de_j\,\mu^i - \mu^j\,\de_j\,c^i = \nn\\
& = \hat{\de}\,c^i + \hat{\hde}\,c^i = \diff \,c^i. 
\end{align}
\begin{align}
(\ell\,s-s\ell)\,\mu^i &= \ell\,s\,\mu^i = \ell\,(-\hat{\hde}\,c^i + c^j\,\de_j\,\mu^i + \mu^j\,\de_j\,c^i) = \nn\\
& = -\hat{\hde}\,\mu^i + (\diff z^j + \mu^j)\,\de_j\,\mu^i + \mu^j\,\de_j\,\mu^j = \nn\\
& = -\hat{\hde}\,\mu^i + \mu^j\,\de_j\,\mu^i + \hat{\de}\,\mu^i + \hat{\hde}\,\mu^i - \hat{\hde}\,\mu^i + \mu^j\,\de_j\,\mu^j = \nn\\
& = - 2\,F^i + \diff\,\mu^i = \diff\,\mu^i.
\end{align}
These properties allow to show that\footnote{To prove it, use $e^{-\ell}\,\delta\,e^\ell = e^{-[\ell,\cdot]}\,\delta$ and expand the exponential in power series.}
\begin{equation}
e^{-\ell}\,\delta\,e^\ell = s,
\end{equation}
which defines the isomorphism between the $\delta$-cohomology and the $s$-cohomology: Thanks to this identity, if $\omega^\natural$ is an $s$-cocycle, then $e^{\ell}\,\omega^\natural$ is a $\delta$-cocycle.\footnote{See Appendix \ref{AppendixA} for some general facts on $\ell$ and on the descent and references.}

Remarkably, 
\begin{equation}
\mathcal{M}^i = e^{\ell}\,c^i,
\end{equation}
since the exponential is truncated at the first order because $c^i$ has ghost number one. This relation is very useful, because it allows to select the component of a given ghost number in a polyform out of the $\mathcal{M}$'s in a simple way. For example,
\begin{align}
\mathcal{M}^i\,\mathcal{M}^j &= (e^\ell\,c^i)\,(e^\ell\,c^j) = e^\ell\,(c^i\,c^j) = c^i\,c^j + \ell\,(c^i\,c^j) + \tfrac{1}{2}\,\ell^2\,(c^i\,c^j),
\end{align}
where in the second step we use the Leibniz identity for the exponential of a derivative.\footnote{If $\de\,(a\,b)=\de\,a\,b+a\,\de\,b$, then $e^{\de}\,(a\,b)=(e^{\de}\,a)\,(e^{\de}\,b)$.} The first, the second and the third terms are the components of ghost number two, one and zero respectively. Namely, the last two ones read 
\begin{equations}
\ell\,(c^i\,c^j) =&\, (\diff z^i + \mu^i)\,\,c^j + c^i\,(\diff z^j + \mu^j),\\
\tfrac{1}{2}\,\ell^2\,(c^i\,c^j) =&\, (\diff z^i + \mu^i)\,(\diff z^j + \mu^j).
\end{equations}

The isomorphism between the $\delta$-cohomology and the $s$-cohomology allows to show that the Dolbeault operator $\hat{\hde}$ can be excluded in computing the $\delta$-cohomology. 

Consider the $s$-cohomology on zero-form polynomials of a given ghost number, generated by $\{\diff z^i, \diff z^{\hi}, \de_i, \de_{\hi}, \mu^i_{\hi}, c^i\}$  in $2\,n$ dimensions. Define the following filtration of \textsc{brst} rules \eqref{BeltramiBRSTRules1}--\eqref{BeltramiBRSTRules2}: \cite{Imbimbo:Unpublished1}
\begin{equations}
& s = s_0 + s_1,\\
& s_0\,c^i = 0, \quad s_0\,\mu^i_{\hi} = \de_{\hi}\,c^i,\\
& s_1\,c^i = c^j\,\de_j\,c^i, \quad s_1\,\mu^i_{\hi} =  c^j\,\de_j\,\mu^i_{\hi}-\de_j\,c^i\,\mu^j_{\hi},
\end{equations} 
As usual in filtration, $n=0,1$ in $s_n$ is the degree in the number of fields or ghosts: $s_n$ increases the number of fields or ghosts by $n$.
Notice that $(\de_{i_1}\dots\de_{i_n}\,\hde_{j_1}\dots\hde_{j_m}\,\mu^i_{\hi};$ $\de_{i_1}\dots\de_{i_n}\,\hde_{j_1}\dots\hde_{j_m}\,\de_{\hi}\,c^i)$ is a $s_0$-doublet, for each $n,m \in \mathbb{N}$. By the doublet theorem \cite{Brandt:1989gv, Brandt:1989gy, Brandt:1989rd}, it is pulled out from the $s_0$-cohomology. But, according to the standard filtration theorem \cite{Brandt:1989gv, Brandt:1989gy, Brandt:1989rd}, the $s$-cohomology is contained in the $s_0$-cohomology.
This implies that the previous doublets are also pulled out from the $s$-cohomology. So, the $s$-cohomology on zero-forms is generated by $\de_{i_1}\dots\de_{i_n}\,c^i$, the antiholomorphic derivatives being excluded. Therefore, the $\delta$-cohomology is generated by $\de_{i_1}\dots\de_{i_n}\,\mathcal{M}^i$.

Now, let us find $\delta$-cocycles using the Stora-Zumino method. Following the strategy adopted in the two-dimensional case, we can introduce the analogous of the $\mathcal{R}$ ``curvature":
\begin{equation}
\mathcal{R}_i{}^j = \mathcal{M}^k\,\de_k\,\de_i\,\mathcal{M}^j.
\end{equation}
Indeed, taking the derivative of $\delta\,\mathcal{M}^j$
\begin{equation}
\delta\,\de_i\,\mathcal{M}^j = \mathcal{R}_i{}^j +  \de_i\,\mathcal{M}^k\,\de_k\,\mathcal{M}^j,
\end{equation}
or equivalently
\begin{equation}
\delta\,\de\,\mathcal{M} = \mathcal{R} + (\de\,\mathcal{M})^2,
\end{equation}
we see that $\mathcal{R}_i{}^j$ is the ``curvature" of the ``connection" $\de_i\,\mathcal{M}^j$. Notice that, if $n>1$, the connection is non-abelian. As usual, $\mathcal{R}$ obeys to a Bianchi identity:
\begin{equation}
\delta\,\mathcal{R} = [\de\,\mathcal{M},\mathcal{R}].
\end{equation}
$\mathcal{R}_i{}^j$ fails to be horizontal:
\begin{align}
\mathcal{R}_i{}^j &= \diff z^k\,\de_k\,\de_i\,\mu^j + \mu^k\,\de_k\,\de_i\,\mu^j \,+\nn\\
& + \diff z^k\,\de_k\,\de_i\,c^j + \mu^k\,\de_k\,\de_i\,c^j + c^k\,\de_k\,\de_i\,\mu^j \,+\nn\\
& + c^k\,\de_k\,\de_i\,c^j,
\end{align}
so that it is an instance of \eqref{F2ghosts}. Nevertheless, one can produce vanishing invariants in $2\,n$ dimensions, by considering traces of powers of $\mathcal{R}$. In particular, if we want to find anomalies in $2\,n$ dimensions, we have to produce a $\delta$-cocycle of total degree $2\,n+1$. So, according to the Stora-Zumino method, we need vanishing invariants of total degree $2\,n+2$. Since $\mathcal{R}$ has total degree two, we consider traces of powers of $n+1$ $\mathcal{R}$'s.

Let us consider for definiteness the case $n=2$ (four dimensions). At the end, we will be able to generalise to arbitrary $n$. We have to write down invariants with three $\mathcal{R}$'s. There are three possibilities:
\begin{equation}
P_{111} = (\Tr \mathcal{R})^3, \quad P_{12} = \Tr \mathcal{R}\,\Tr \mathcal{R}^2, \quad P_3 = \Tr \mathcal{R}^3.
\end{equation}
They are classified by the partitions $1+1+1 = 1+2 = 3$, which dictate how to take the trace. To see that they vanish \emph{in four dimensions}, notice that each of them involves three $\mathcal{M}^i$ free of derivatives, but the index $i$ takes only two values in four dimensions, so that in the combination $\mathcal{M}^i\,\mathcal{M}^j\,\mathcal{M}^k$ one of the possible values of the index must be repeated. Thus, there is twice the same $\mathcal{M}$ in that combination. Therefore, it vanishes for anticommutativity:
\begin{equation}
\mathcal{M}^i\,\mathcal{M}^j\,\mathcal{M}^k = 0, \;\; \text{if}\;\; n=2.
\end{equation}
Analogously, any possible combination of five (or more) $\de_i\,\mathcal{M}^j$ vanishes in four dimensions, since there are four possible matrices $\de_i\,\mathcal{M}^j$ as the indices $i,j=1,2$ vary:
\begin{equation}\label{5Gammas}
\de_{i_1}\,\mathcal{M}^{j_1}\,\de_{i_2}\,\mathcal{M}^{j_2}\,\de_{i_3}\,\mathcal{M}^{j_3}\,\de_{i_4}\,\mathcal{M}^{j_4}\,\de_{i_5}\,\mathcal{M}^{j_5} = 0, \;\; \text{if}\;\; n=2. 
\end{equation}
These conclusions can be also confirmed by expanding the polyforms in their components and using the same logic. For instance,
\begin{align}
\mathcal{M}^i\,\mathcal{M}^j\,\mathcal{M}^k &= c^i\,c^j\,c^k + \ell\,(c^i\,c^j\,c^k) + \tfrac{1}{2}\,\ell^2\,(c^i\,c^j\,c^k) + \tfrac{1}{6}\,\ell^3(c^i\,c^j\,c^k) = \nn\\
& = c^i\,c^j\,c^k \,+ \nn\\
& + c^i\,c^j\,\diff z^k + c^i\,\diff z^j\,c^k + \diff z^i\,c^j\,c^k \,+\nn\\
& + c^i\,c^j\,\mu^k  + c^i\,\mu^j\,c^k + \mu^i\,c^j\,c^k \,+\nn\\
& + \diff z^i \,\diff z^j \,c^k + \diff z^i\,c^j\,\diff z^k + c^i\,\diff z^j\,\diff z^k \,+\nn\\
& + \diff z^i \,\mu^j \,c^k + \mu^i\,\diff z^j \,c^k + \diff z^i\,c^j\,\mu^k + \mu^i\,c^j\,\diff z^k+ c^i\,\diff z^j \,\mu^k + c^i\,\mu^j\,\diff z^k \,+\nn\\
& + \mu^i\,\mu^j\,c^k + \mu^i\,c^j\,\mu^j + c^i \,\mu^j\,\mu^k \,+\nn\\
& + \diff z^i\,\diff z^j\,\diff z^k \,+ \nn\\
& + \diff z^i\,\diff z^j\,\mu^k + \diff z^i\,\mu^j\,\diff z^k + \mu^i\,\diff z^j\,\diff z^k \,+\nn\\
& + \diff z^i\,\mu^j\,\mu^k + \mu^i\,\diff z^j\,\mu^k + \mu^i \,\mu^j \,\diff z^k \,+\nn\\
& + \mu^i\,\mu^j\,\mu^k = 0,
\end{align}
in which each line separately vanishes, for any choice of the indices $i,j,k=1,2$.

The invariants we listed before can be all written as $\delta$-coboundaries:
\begin{equation}
\Tr \mathcal{R} = \delta\,Q_1, \quad
\Tr \mathcal{R}^2 = \delta\,Q_3, \quad
\Tr \mathcal{R}^3 = \delta\,Q_5,
\end{equation}
by means of the Chern polynomials
\begin{equations}
& Q_1 = \Tr(\de\,\mathcal{M}),\\
& Q_3 = \Tr(\de\,\mathcal{M}\,\mathcal{R} + \tfrac{1}{3}\,(\de\,\mathcal{M})^3),\\
& Q_5 = \Tr(\de\,\mathcal{M}\,\mathcal{R}^2 +\tfrac{1}{2}\,(\de\,\mathcal{M})^3\,R + \tfrac{1}{10}\,(\de\,\mathcal{M})^5).
\end{equations}
Actually, the last term in $Q_5$ vanishes in four dimensions, thanks to \eqref{5Gammas}. Therefore, we can write:
\begin{equations}
0 &= P_{111} = (\Tr \mathcal{R})^3 = \delta\,\Tr(\de\,\mathcal{M})\,(\Tr \mathcal{R})^2 = \delta\,(\Tr(\de\,\mathcal{M})\,(\Tr \mathcal{R})^2),\label{P111}\\
0 &= P_{12} = \Tr \mathcal{R}\,\Tr \mathcal{R}^2 = \delta\,\Tr(\de\,\mathcal{M})\,\Tr \mathcal{R}^2 = \delta\,(\Tr(\de\,\mathcal{M})\,\Tr \mathcal{R}^2),\label{P12}\\
0 &= P_3 = \Tr \mathcal{R}^3 = \delta\,Q_5 = \delta\,\Tr(\de\,\mathcal{M}\,\mathcal{R}^2 +\tfrac{1}{2}\,(\de\,\mathcal{M})^3\,\mathcal{R}).\label{P3}
\end{equations}
So, we have produced the following $\delta$-cocycles without any computation:
\begin{equations}
\omega_{111} &= \Tr(\de\,\mathcal{M})\,(\Tr \mathcal{R})^2,\\
\omega_{12} &= \Tr(\de\,\mathcal{M})\,\Tr \mathcal{R}^2,\\
\omega_3 &= \Tr(\de\,\mathcal{M}\,\mathcal{R}^2 + \tfrac{1}{2}\,(\de\,\mathcal{M})^3\,\mathcal{R}).
\end{equations}
The corresponding components of ghost number one are
\begin{equations}
(\omega_{111})^{(4)}_1 &= \Tr(\de\,c)\,\Tr(\hat{\de}\,(\de\,\mu))\,\Tr(\hat{\de}\,(\de\,\mu)) + 2\,\Tr(\de\,\mu)\,\Tr(\hat{\de}\,(\de\,c))\,\Tr(\hat{\de}\,(\de\,\mu)),\\
(\omega_{12})^{(4)}_1 &=  \Tr(\de\,c)\,\text{Tr}\,(\hat{\de}\,(\de\,\mu)\,\hat{\de}\,(\de\,\mu)) + 2\,\Tr(\de\,\mu)\,\text{Tr}\,(\hat{\de}\,(\de\,c)\,\hat{\de}\,(\de\,\mu)), \\
(\omega_{3})^{(4)}_1 &= \Tr(\de\,c\,\hat{\de}\,(\de\,\mu)\,\hat{\de}\,(\de\,\mu)) + \Tr(\de\,\mu\,\hat{\de}\,(\de\,c)\,\hat{\de}\,(\de\,\mu)) + \Tr(\de\,\mu\,\hat{\de}\,(\de\,\mu)\,\hat{\de}\,(\de\,c)),
\end{equations}
where $\de\,c$ and $\de\,\mu$ are the matrices with components $\de_i\,c^j$ and $\de_i\,\mu^j$ and $\hat{\de}$ is the operator $\diff z^i\,\de_i$ acting on $\de\,c$ and $\de\,\mu$. Notice that
\begin{equation}
(\de\,\mu)\,\hat{\de}\,(\de\,c)\,\hat{\de}\,(\de\,\mu) = 
(\de\,\mu)\,(\hat{\de} + \hat{\hde})\,(\de\,\mu)\,(\hat{\de} + \hat{\hde})\,(\de\,c) = (\de\,\mu)\,\diff\,(\de\,\mu)\,\diff\,(\de\,c).\label{Trick}
\end{equation}
Indeed, the terms added in the first step vanish, since they contain more than three $\diff z^{\hi}$ (remember that $\mu^i$ is a $(0,1)$-form). Therefore, using 
\begin{equation}
(\de\,\mu)\,\diff\,(\de\,c)\,\diff(\de\,\mu) = 
\diff\,(\de\,\mu)(\de\,c)\,\diff\,(\de\,\mu) - \diff\,(\de\,\mu\,\de\,c\,\diff\,\de\,\mu),
\end{equation}
and so on, we can write
\begin{equations}
(\omega_{111})^{(4)}_1 &\simeq 3\,\Tr(\de\,c)\,\Tr(\diff\,\de\,\mu)\,\Tr(\diff\,\de\,\mu),\label{cocycle4d1}\\ 
(\omega_{12})^{(4)}_1 &\simeq \Tr(\de\,c)\,\Tr(\diff\,\de\,\mu\,\diff\,\de\,\mu) + 2\,\Tr(\diff\,\de\,\mu)\,\Tr(\de\,c\,\diff\,\de\,\mu),\label{cocycle4d2}\\
(\omega_{3})^{(4)}_1 &\simeq 3\,\Tr(\de\,c\,\diff\,\de\,\mu\,\diff\,\de\,\mu),\label{cocycle4d3}
\end{equations}
which generalise in four dimensions the Gelfand-Fuchs anomaly \eqref{GF_Anomaly} (which is $\propto \de\,c\,\diff\,\de\,\mu$, using the notation of this section, since $\mu = \diff \hz\,\mu_{\hz}^z$ and $\diff = \diff\,z\,\de_z + \diff \hz\,\de_{\hz}$).

It is interesting to note that, when we wrote $P_{12}$ as a $\delta$-coboundary, we could have written $P_{12} = \delta\,(\Tr \mathcal{R}\,Q_3)$ instead of $P_{12} = \delta\,(Q_1\,\Tr \mathcal{R}^2)$, but the solution $\Tr \mathcal{R}\,Q_3$ is equivalent to $Q_1\,\Tr \mathcal{R}^2$, since they differ by a $\delta$-exact term:
\begin{equation}
\delta\,(Q_1\,Q_3) = \Tr \mathcal{R}\,Q_3 - Q_1\,\Tr \mathcal{R}^2.
\end{equation}

Let us extend to arbitrary $2\,n$ dimensions. The invariants involving $n+1$ $\mathcal{R}$'s are classified by the partitions of $n+1$:
\begin{equations}
P_{1\dots 1} &= (\Tr \mathcal{R})^{n+1}, \\
P_{1\dots 12} &= (\Tr \mathcal{R})^{n-1}\,\Tr \mathcal{R}^2, \\
P_{1\dots 122} &= (\Tr \mathcal{R})^{n-3}\,(\Tr \mathcal{R}^2)^2,\\
&\vdots\nn\\
P_{n+1} &= \Tr \mathcal{R}^{n+1}.
\end{equations}
The arbitrary invariant is
\begin{equation}
P_{k_1\dots k_p} = \Tr \mathcal{R}^{k_1}\dots \Tr \mathcal{R}^{k_{p}},
\end{equation}
corresponding to the partition $k_1 + \dots + k_{p} = n+1$, with $1 \leqslant k_1 \leqslant \dots \leqslant k_{p} \leqslant n+1$, $p$ being the number of (possibly repeated) integers in the partition. On the one hand, 
\begin{equation}
P_{k_1\dots k_p} = 0, \quad \text{in}\;2\,n\;\text{dimensions},
\end{equation}
since it involves $n+1$ $\mathcal{M}$'s and we have
\begin{equation}
\mathcal{M}^{i_1}\dots \mathcal{M}^{i_{n+1}} = 0\;\;\;\text{in}\;2\,n\;\text{dimensions.}
\end{equation}
On the other hand, using
\begin{equation}
\Tr \mathcal{R}^k = \delta\,Q_{2k-1}(\de\,\mathcal{M},\mathcal{R}),
\end{equation}
where $Q_{2k-1}(\de\,\mathcal{M},\mathcal{R})$ is the Chern polynomial of degree $2\,k-1$, generalisation of \eqref{P111}-\eqref{P3} yields
\begin{equation}
P_{k_1\dots k_p} = \delta\,(Q_{2k_1-1}\,\Tr \mathcal{R}^{k_2}\dots\Tr \mathcal{R}^{k_p}).
\end{equation}
Therefore,
\begin{equation}
\delta\,(Q_{2k_1-1}\,\Tr \mathcal{R}^{k_2}\dots\Tr \mathcal{R}^{k_p}) = 0,
\end{equation}
which is the Stora-Zumino equation  \eqref{SZeq} or \eqref{SZeqBis} for the following polyform of total degree $2\,n+1$
\begin{equation}
\omega_{k_1k_2\dots k_p} = Q_{2k_1-1}\,\Tr \mathcal{R}^{k_2}\dots\Tr \mathcal{R}^{k_p},
\end{equation}
whose component $(\omega_{k_1k_2\dots k_p})_1^{(2n)}$ of ghost number 1 and form degree $2\,n$ is then an anomaly. 

Notice that we have chosen to express $P_{k_1\dots k_p}$ as a $\delta$-coboundary by making $Q_{2k_1-1}$ explicit, but the possible other ways, which make $Q_{2k_l-1}$ explicit, for any $1\leqslant l \leqslant p$, are all equivalent, because they differ by a $\delta$ exact term. Indeed, setting $1 \leqslant l < m \leqslant n+1$ and denoting the omission with a caret,
\begin{equations}
\delta\,(&\,Q_{2k_l-1}\,Q_{2k_m-1}\,\Tr \mathcal{R}^{k_1}\dots\hat{\Tr \mathcal{R}^{k_l}}\dots\hat{\Tr \mathcal{R}^{k_m}}\dots \Tr \mathcal{R}^{k_p}) = \nn\\
& = Q_{2k_m-1}\,\Tr \mathcal{R}^{k_1}\dots\hat{\Tr \mathcal{R}^{k_m}}\dots \Tr \mathcal{R}^{k_p} - Q_{2k_l-1}\,\Tr \mathcal{R}^{k_1}\dots\hat{\Tr \mathcal{R}^{k_l}}\dots \Tr \mathcal{R}^{k_p}.
\end{equations}

Let us find a closed formula for the Kodaira-Spencer anomaly. In the $n=2$ case, after having integrating by parts, the anomaly was made up of $\de\,c\,\diff\,\de\,\mu\,\diff\,\de\,\mu$, traced in all the possible ways, according to the partitions of $n+1=3$. In general, fixed an arbitrary partition of $n+1$ as before, there are $k_l$ possible dispositions for $\de\,c$ in each trace with $k_l$ terms; the remaining $k_l-1$ places are occupied by $\diff\,\de\,\mu$ in the same trace; all the other traces are filled by $\diff\,\de\,\mu$. Therefore, the result is
\begin{equation}\label{FinalResult}
(\omega_{k_1k_2\dots k_p})_1^{(2n)} \simeq
\sum_{l=1}^{n+1} k_l\,\Tr(\diff\,\de\,\mu)^{k_1}\dots \Tr(\de\,c\,(\diff\,\de\,\mu)^{k_l-1})\dots\Tr(\diff\,\de\,\mu)^{k_p}.
\end{equation}
For example, in six dimensions ($n=3$), we have:
\begin{equations}
(\omega_{1111})_1^{(6)} &\simeq 4\,\Tr(\de\,c)\,(\Tr(\diff\,\de\,\mu))^{3},\\
(\omega_{112})_1^{(6)} &\simeq 2\,\Tr(\de\,c)\,\Tr(\diff\,\de\,\mu)\Tr(\diff\,\de\,\mu)^2 + 2\,\Tr(\de\,c\,\diff\,\de\,\mu)\,(\Tr(\diff\,\de\,\mu))^2,\\
(\omega_{22})_1^{(6)} &\simeq 4\,\Tr(\de\,c\,\diff\,\de\,\mu)\,\Tr(\diff\,\de\,\mu)^2,\\
(\omega_{13})_1^{(6)} &\simeq \Tr(\de\,c)\,\Tr(\diff\,\de\,\mu)^3 + 3\,\Tr(\de\,c\,(\diff\,\de\,\mu)^2)\,\Tr(\diff\,\de\,\mu),\\
(\omega_{4})_1^{(6)} &\simeq 4\,\Tr(\de\,c\,(\diff\,\de\,\mu)^3).
\end{equations}

\section{Conclusions}

We computed the Kodaira-Spencer anomalies in the \textsc{brst} formulation of $2\,n$-dimensional gravitational theories in Beltrami parametrisation, using an extension of the Stora-Zumino method. This is an example of application of the Stora-Zumino method in a context in which the breaking of the horizontality condition of the polycurvature is maximal: The polycurvature has components of ghost number one and two, as in \eqref{F2ghosts}, generalising the previously known cases, in which the polycurvature is horizontal \eqref{F0ghost}, or it has only a non-vanishing ghost number one component \eqref{F1ghost}.

The result of the computation is the equation \eqref{FinalResult}. The Stora-Zumino approach simplifies significantly the analysis in \cite{Bandelloni:1998vp}. The anomalies we found have a clear geometric interpretation, since they are expressed in terms of Chern polynomials and Pontryagin invariants. The method was able to capture all the solutions found in equations (5.52) and (5.53) of \cite{Bandelloni:1998vp}, but many other possibilities were obtained, since our solutions are classified by the partitions of $n+1$, instead of those of $n$. Namely, the anomalies obtained in \cite{Bandelloni:1998vp} are those which are related, in our approach, to the invariants containing a factor $\Tr \mathcal{R}$. In other terms, only the anomalies $(\omega_{1k_2\dots k_p})_1^{(2n)}$ are considered. In particular, the anomaly corresponding to the partition $k_l = 0$, $l<p$,  $k_p = n+1$, that is
\begin{equation}
(\omega_{n+1})^{(2n)}_1 \simeq \Tr(\de\,c\,(\diff\,\de\,\mu)^n),
\end{equation}
which is not captured in \cite{Bandelloni:1998vp}, corresponds to the one proposed in equation (1.11) in \cite{Losev:1996up}. So, we proved that this anomaly is a consistent anomaly in the \textsc{brst} sense too.

\section*{Acknowledgments}

The author is deeply indebted to Camillo Imbimbo for stimulating discussions and illuminating comments.

\appendix
\section{Homotopy operator and polyforms}\label{AppendixA}

A topological theory is characterised by the existence of an homotopy operator $\ell$, which allows to state that the $s$-cohomology and the $\delta$-cohomology are isomorph. But the $\delta$-cohomology is equivalent to the $s$-cohomology modulo $\diff$. This shows immediately that the forms $\omega^{(p)}_{d+1-p}$ involved in the $d$-dimensional descent can be obtained knowing the elements ${\omega^{\natural}}^{(p)}_{d+1-p}$ in the $s$-cohomology on $p$-form of ghost number $d+1-p$ \cite{Sorella:1992dr, Sorella:1993kq, Blaga:1995rv, Tataru:1995ru}.

Let us perform the direct computation, starting from the descent (for definiteness in two dimensions):
\begin{equations}
& s\,\omega_1^{(2)} = -\diff \omega_2^{(1)},\label{topeq}\\
& s\,\omega_2^{(1)} = -\diff \omega_3^{(0)},\label{sdeq}\\
& s\,\omega_3^{(0)} = 0
\end{equations}
Denote the element in the cohomology with $\natural$. Arbitrary forms are denoted by $\sim$. The most general solution of the bottom equation is
\begin{equation}
\omega_3^{(0)} = {\omega^\natural}_3^{(0)} + s\,\tilde{\omega}_2^{(0)}.
\end{equation}
Compute the exterior differential:
\begin{align}
\diff\omega_3^{(0)} &= (\ell\,s-s\,\ell)\,\omega_3^{(0)} = -s\,\ell\,\omega_3^{(0)} = -s\,\ell\,{\omega^\natural}_3^{(0)} - s\,\ell\,s\,\tilde{\omega}_2^{(0)} =\nn\\
& = -s\,\ell\,{\omega^\natural}_3^{(0)} - s\,(s\,\ell + \diff)\,\tilde{\omega}_2^{(0)}  = -s\,\ell\,{\omega^\natural}_3^{(0)} - s\,\diff\,\tilde{\omega}_2^{(0)}.
\end{align}
Plugging into the second equation \eqref{sdeq},
\begin{equation}
s\,(\omega_2^{(1)} - \ell\,{\omega^\natural}_3^{(0)} - \diff \,\tilde{\omega}_2^{(0)}) = 0,
\end{equation}
whose general solution is
\begin{equation}
\omega_2^{(1)} = {\omega^\natural}_2^{(1)} + \ell\,{\omega^\natural}_3^{(0)} + \diff\,\tilde{\omega}_2^{(0)} + s\,\tilde{\omega}_1^{(1)}.
\end{equation}
Compute its exterior differential:
\begin{equation}
\diff\,\omega_2^{(1)} = (\ell\,s - s\,\ell)\,{\omega^\natural}_2^{(1)} + \diff\,\ell\,{\omega^\natural}_3^{(0)} - s\,\diff\tilde{\omega}_1^{(1)}.
\end{equation}
Notice that
\begin{align}
\diff\,\ell\,{\omega^\natural}_3^{(0)} &= (\ell\,s-s\,\ell)\,\ell\,{\omega^\natural}_3^{(0)} = 
\ell\,s\,\ell\,{\omega^\natural}_3^{(0)} - s\,\ell^2\,{\omega^\natural}_3^{(0)} = \nn\\
&= \ell\,(\ell\,s-\diff)\,{\omega^\natural}_3^{(0)} - s\,\ell^2\,{\omega^\natural}_3^{(0)} = -\diff\,\ell\,{\omega^\natural}_3^{(0)}-s\,\ell^2\,{\omega^\natural}_3^{(0)},
\end{align}
so that
\begin{equation}
\diff\,\ell\,{\omega^\natural}_3^{(0)} = -\tfrac{1}{2}\,s\,\ell^2\,{\omega^\natural}_3^{(0)}.
\end{equation}
Therefore, we get
\begin{equation}
\diff\,\omega_2^{(1)} = - s\,(\ell\,{\omega^\natural}_2^{(1)} + \tfrac{1}{2}\,\ell^2\,{\omega^\natural}_3^{(0)} + \diff\,\tilde{\omega}_1^{(2)}).
\end{equation}
Plugging into the top equation \eqref{topeq},
\begin{equation}
s\,(\omega_1^{(2)} - \ell\,{\omega^\natural}_2^{(1)} - \tfrac{1}{2}\,\ell^2\,{\omega^\natural}_3^{(0)} - \diff\,\tilde{\omega}_1^{(2)}) = 0,
\end{equation}
whose general solution is
\begin{equation}
\omega_1^{(2)} = {\omega^\natural}_1^{(2)}  + \ell\,{\omega^\natural}_2^{(1)} + \tfrac{1}{2}\,\ell^2\,{\omega^\natural}_3^{(0)} + \diff\,\tilde{\omega}_1^{(2)} + s\,\tilde{\omega}_0^{(2)}.
\end{equation}
Summarising, the solution is
\begin{equations}
\omega_1^{(2)} &= {\omega^\natural}_1^{(2)}  + \ell\,{\omega^\natural}_2^{(1)} + \tfrac{1}{2}\,\ell^2\,{\omega^\natural}_3^{(0)} + \diff\,\tilde{\omega}_1^{(2)} + s\,\tilde{\omega}_0^{(2)},\\
\omega_2^{(1)} &= {\omega^\natural}_2^{(1)} + \ell\,{\omega^\natural}_3^{(0)} + \diff\,\tilde{\omega}_2^{(0)} + s\,\tilde{\omega}_1^{(1)},\\
\omega_3^{(0)} &= {\omega^\natural}_3^{(0)} + s\,\tilde{\omega}_2^{(0)}.
\end{equations}

This result can be readly generalised to $d$ dimensions: 
\begin{align}
\omega^{(p)}_{d+1-p} &= {\omega^{\natural}}^{(p)}_{d+1-p} + \ell\,{\omega^{\natural}}^{(p-1)}_{d+2-p} + \tfrac{1}{2}\,\ell^2\,{\omega^{\natural}}^{(p-2)}_{d+3-p} + \dots \,+\nn\\
& + \tfrac{1}{p!}\,\ell^{p}\,{\omega^{\natural}}^{(0)}_{d+1} + \diff \tilde{\omega}_{d+1-p}^{(p-1)} + s\,\tilde{\omega}_{d-p}^{(p)},
\end{align}
for any $0 \leqslant p \leqslant d$. 

The whole solution fits nicely into a polyform, as $p$ varies:
\begin{align}
\omega^{(d)}_1 + \dots + \omega^{(0)}_{d+1} &=  (1 + \ell + \tfrac{1}{2}\,\ell^2 + \dots + \tfrac{1}{d!}\,\ell^d)\,{\omega^{\natural}}^{(0)}_{d+1} \,+\nn\\
& + (1 + \ell + \tfrac{1}{2}\,\ell^2 + \dots + \tfrac{1}{(d-1)!}\,\ell^{d-1})\,{\omega^{\natural}}^{(1)}_{d} +\dots \,+\nn\\
& + {\omega^{\natural}}^{(d)}_1 + (\diff + s)\,(\tilde{\omega}^{(d)}_0 + \dots + \tilde{\omega}^{(0)}_d),
\end{align}
that is,
\begin{equation}
\omega_{d+1} = e^{\ell}\,\omega^{\natural}_{d+1} + \delta\,\tilde{\omega}_d,
\end{equation}
having defined
\begin{equations}
\delta &:= \diff + s,\\
\omega_{d+1} &:= \omega_1^{(d)} + \dots + \omega_{d+1}^{(0)},\\
\omega^{\natural}_{d+1} &:= {\omega^{\natural}}^{(d)}_1 + \dots + {\omega^{\natural}}^{(0)}_{d+1},\\
\tilde{\omega}_d &:= \tilde{\omega}^{(d)}_0 + \dots + \tilde{\omega}^{(0)}_d.
\end{equations}
This is in agreement with the isomorphism between $s$- and $\delta$-cohomology: if $\omega^{\natural}_{d+1}$ is an $s$-cocycle, then $e^{\ell}\,\omega^{\natural}_{d+1}$ is a $\delta$-cocycle.

\section{Direct computation of 4d $\delta$-cocycles}

To confirm the result obtained using the Stora-Zumino method in computing $\delta$-cocycles of total degree five in four dimensions, it is instructive to perform a direct computation. Consider the most general polyform of degree five $\omega_5$, generated by $\de_i\,\mathcal{M}^j$ and $\mathcal{R}_i{}^j$. We have eight possible terms, the first one vanishing in four dimensions:
\begin{equations}
\omega_{(0)} &= \Tr(\de\,\mathcal{M})^5 = 0,\quad \text{if}\;n=2,\\
\omega_{(1)} &= \Tr(\de\,\mathcal{M})\,\Tr \mathcal{R}^2,\\
\omega_{(2)} &= \Tr(\de\,\mathcal{M}\,\mathcal{R}^2),\\
\omega_{(3)} &= \Tr(\de\,\mathcal{M})^3\,\Tr \mathcal{R},\\
\omega_{(4)} &= \Tr((\de\,\mathcal{M})^3\,\mathcal{R}),\\
\omega_{(5)} &= \Tr(\de\,\mathcal{M})\,(\Tr \mathcal{R})^2,\\
\omega_{(6)} &= \Tr(\de\,\mathcal{M}\,\mathcal{R})\,\Tr \mathcal{R},\\
\omega_{(7)} &= \Tr(\de\,\mathcal{M})\,\Tr((\de\,\mathcal{M})^2\,\mathcal{R}),
\end{equations}
so that
\begin{equation}
\omega_5 = \sum_{k=1}^7 x_k\,\omega_{(k)},
\end{equation}
for any coefficients $x_k$. Now compute the $\delta$-variation of $\omega_5$ and impose it to vanish. One finds:
\begin{equation}
\delta\,\omega_5 = 0 \Leftrightarrow x_4 = \tfrac{1}{2}\,x_2, \;\; x_7 = x_6 -3\,x_3.
\end{equation}
Therefore, $\omega_5$ is a $\delta$-cocycle if and only if
\begin{equation}
\omega_5 = x_1\,\omega_{(1)} + x_2\,(\omega_{(2)} + \tfrac{1}{2}\,\omega_{(4)}) -3\,x_3\,(\omega_{(7)}-\tfrac{1}{3}\,\omega_{(3)}) + x_5\,\omega_{(5)} + x_6\,(\omega_{(6)} + \omega_{(7)}),
\end{equation}
In particular, notice that the second solution corresponds to the five-dimensional Chern polynomial $Q_5$:
\begin{equation}
Q_5 = \omega_{(2)} + \tfrac{1}{2}\,\omega_{(4)} + \tfrac{1}{10}\,\omega_{(0)},
\end{equation} 
where the last term vanishes in four dimensions. 

Now, we have to understand if the previous solutions are all independent or not, and if some of these solutions are $\delta$-exact (to be excluded). Consider the most general polyform $\psi_4$ of total degree four generated by $\de_i\,\mathcal{M}^j$ and $\mathcal{R}_i{}^j$. There are five possible terms:
\begin{equations}
\psi_{(1)} &= \Tr \mathcal{R}^2,\\
\psi_{(2)} &= (\Tr \mathcal{R})^2,\\
\psi_{(3)} &= \Tr((\de\,\mathcal{M})^2\,\mathcal{R}),\\
\psi_{(4)} &= \Tr(\de\,\mathcal{M})\,\Tr(\de\,\mathcal{M}\,\mathcal{R}),\\
\psi_{(5)} &= \Tr(\de\,\mathcal{M})\,\Tr(\de\,\mathcal{M})^3.
\end{equations}
The first three $\psi_{(i)}$ are $\delta$-cocycles:
\begin{equation}
\delta\,\psi_{(i)} = 0, \;\;i=1,2,3.
\end{equation}
The remaining ones can be used to find relations between the previous solutions. In particular, one can verify that
\begin{equations}
&\omega_{(6)} + \omega_{(7)} = \omega_{(1)} + \delta\,\psi_{(4)}, \\
&\omega_{(7)} - \tfrac{1}{3}\,\omega_{(3)} = -\tfrac{1}{3}\,\delta\,\psi_{(5)},
\end{equations}
showing that $\omega_{(6)} + \omega_{(7)}$ is equivalent to $\omega_{(1)}$ and $\omega_{(7)} - \frac{1}{3}\,\omega_{(3)}$ is trivial. Therefore, we can drop these solutions, remaining with
\begin{equation}
\omega_5 = x_1\,\omega_{(1)} + x_2\,Q_5 + x_5\,\omega_{(5)}.
\end{equation}
The three survived solutions are respectively the three solutions \eqref{cocycle4d1}, \eqref{cocycle4d2}, and \eqref{cocycle4d3} captured by the Stora-Zumino method.

\section{A counterterm for an anomaly}

The computation of local $\delta$-cohomology on the space of finite polynomials of $c^i$ and $\mu^i$ was performed throughout the paper. Nevertheless, if we extend the space of fields and if we allow non-polynomial or non-finite polynomial expressions, it would be possible in principle to find local counterterms which eliminate some anomalies. As an example, in this appendix it is shown that the Kodaira-Spencer anomaly corresponding to the partition $k_l = 1, l=1,\dots,p$ can be eliminated by means of a logarithmic counterterm if the integrating matrix $\lambda_i{}^I$ is included in the field space. This fact was pointed out in \cite{Bandelloni:1998vp}, but in that context it is a starting point, whereas it seems more natural that it is an arrival one.

The result has an analogue in the general covariant formulation. Taking into account only the diffeomorphisms, the \textsc{brst} variation of the metric is
\begin{equation}\label{CovariantBRSRules}
s\,g_{\mu\nu} = \nabla_\mu\,\xi_\nu+\nabla_\nu\,\xi_\mu.
\end{equation}
where $\xi^\mu$ is the diffeomorphism anticommuting ghost. As shown in \cite{Bonora:1984pz}, the following cocycle \begin{equation}
\de_\mu\,\xi^\mu\,e_{2n}(R),
\end{equation}
where $e_{2n}(R)$ is the $2\,n$-dimensional Euler form, becomes trivial by taking into account the logarithm of the determinant of the metric. Indeed, there exists a one-weighted vector density $K^\mu$, such that \cite{Chern:1944}
\begin{equation}
e_{2n}(R)=-\de_\mu\,K^\mu\,\diff^{2n}x.
\end{equation} 
Using \eqref{CovariantBRSRules}, one can show that
\begin{equation}
s\,\sqrt{|g|} = \de_\mu\,(\xi^\mu\,\sqrt{|g|}).
\end{equation}
Since $K^\mu$ is a one-weighted vector density, its $s$-variation reads
\begin{equation}
s\,K^\mu = \de_\nu\,(\xi^\nu\,K^\mu) - K^\nu\,\de_\nu\,\xi^\mu,
\end{equation}
Now, consider that
\begin{equation}\label{CovariantCase}
s\,(K^\mu\,\de_\mu\,\log{\sqrt{|g|}}) = \de_\mu\,(K^\mu\,\de_\nu\,\xi^\nu) - \de_\mu\,\xi^\mu\,\de_\nu\,K^\nu.
\end{equation}
Therefore, integrating at both sides, one finds an expression whose $s$-variation gives the integrated anomaly:
\begin{equation}
s\int d^{2n}x\,K^\mu\,\de_\mu\,\log{\sqrt{|g|}} = 
\int \de_\mu\,\xi^\mu\,e_{2n}(R).
\end{equation}

Let us proceed similarly for the Kodaira-Spencer anomaly $(\omega_{1\dots 1})^{(2n)}_1$. Observe that
\begin{align}
(\omega_{1\dots 1})^{(2n)}_1 &\simeq \Tr(\de\,c)\,
\Tr(\hat{\de}\,(\de\,\mu))\dots \Tr(\hat{\de}\,(\de\,\mu)) = \nn\\
&= \Tr(\de\,c)\,\hat{\de}\,[\Tr(\de\,\mu)\,\Tr(\hat{\de}\,(\de\,\mu))\dots\Tr(\hat{\de}\,(\de\,\mu))] =: -\diff^n z\,\diff^n \hz\,\Tr(\de\,c)\,\de_i\,K^i.
\end{align}
Taking the determinant of the $s$-variation of the integrating matrix, 
\begin{equation}
s\,\det\lambda = \de_i\,(c^i\,\det\lambda).
\end{equation}
Thus, similarly to \eqref{CovariantCase}, 
\begin{align}
s\,(\diff^n z\,\diff^n \hz\,K^i\,\de_i\,\log{\det\lambda}) &= \de_i\,(\diff^n z\,\diff^n \hz\,K^i\,\Tr(\de\,c)) - \diff^n z\,\diff^n \hz\,\de_j\,c^j\,\de_i\,K^i = \nn\\
& = -\hat{\de}\,[\Tr(\de\,\mu)\,\Tr(\hat{\de}\,(\de\,\mu))\dots\Tr(\hat{\de}\,(\de\,\mu))\,\Tr(\de\,c)] + (\omega_{n+1})^{(2n)}_1 = \nn\\
&= -\diff\,[\Tr(\de\,\mu)\,\Tr(\diff\,(\de\,\mu))\dots\Tr(\diff\,(\de\,\mu))\,\Tr(\de\,c)] + (\omega_{n+1})^{(2n)}_1,
\end{align}
where the last step follows from the same trick as in \eqref{Trick}. Finally, integrating the previous relation,
\begin{equation}
s\int \diff^n z\,\diff^n \hz\,K^i\,\de_i\,\log{\det\lambda} = \int (\omega_{n+1})^{(2n)}_1.
\end{equation}

\bibliographystyle{JHEP}
\bibliography{ir}

\end{document}